\documentclass[a4paper,UKenglish,cleveref, autoref, thm-restate]{lipics-v2021}
\makeatletter
\DOIPrefix{}
\makeatother

\usepackage{mathptmx}                  
\usepackage{amsmath}
\usepackage{comment}
\usepackage{cite}

\usepackage{hyperref}
\hypersetup{
    colorlinks=true,
    linkcolor=blue,      
    urlcolor=blue,
    citecolor=blue,
}

\usepackage{amsfonts}
\usepackage{tabularx}
\usepackage{todonotes}
\usepackage{xspace}
\usepackage{tabularx}
\usepackage{colortbl}
\usepackage{float}
\usepackage{paralist}
\usepackage{caption}
\usepackage{subcaption}
\usepackage{lineno}

\newcommand{\rawstress}{raw stress\xspace}
\newcommand{\Rawstress}{Raw stress\xspace}
\newcommand{\RawStress}{Raw Stress\xspace}
\newcommand{\Rawstressmath}{RS\xspace}
\newcommand{\normalizedstress}{normalized stress\xspace}
\newcommand{\Normalizedstress}{Normalized stress\xspace}
\newcommand{\NormalizedStress}{Normalized Stress\xspace}
\newcommand{\Normalizedstressmath}{NS\xspace}
\newcommand{\neatokamadakawaistress}{Kamada-Kawai stress\xspace}
\newcommand{\Neatokamadakawaistress}{Kamada-Kawai stress\xspace}

\newcommand{\minoptstress}{scale-normalized stress\xspace}
\newcommand{\Minoptstress}{Scale-normalized stress\xspace}
\newcommand{\MinoptStress}{Scale-normalized Stress\xspace}
\newcommand{\Minoptstressmath}{SNS\xspace}
\newcommand{\shepardgoodness}{Shepard goodness score\xspace}
\newcommand{\ShepardGoodness}{Shepard Goodness Score\xspace}
\newcommand{\Shepardgoodnessmath}{SGS\xspace}

\newcommand{\kruskalstress}{non-metric stress\xspace}
\newcommand{\Kruskalstress}{Non-metric stress\xspace}
\newcommand{\KruskalStress}{Non-metric Stress\xspace}
\newcommand{\Kruskalstressmath}{NMS\xspace}

\newcommand{\alphaintnorm}{\alpha_{intersection}}
\newcommand{\alphaint}{\alpha_{intersection}}
\newcommand{\alphaminraw}{\alpha_{min}}
\newcommand{\alphaminnorm}{\alpha_{min}}
\newcommand{\alphamin}{\alpha_{min}}

\title{Size Should Not Matter: Evaluating Network Visualizations with Stress} 

\titlerunning{Size Should Not Matter: Evaluating Network Visualizations with Stress}

\author{Reyan Ahmed}
{University of Arizona }
{abureyanahmed@arizona.edu}
{0000-0001-6830-9053}
{}

\author{Cesim Erten}
{University of Arizona}
{cesim@arizona.edu}
{}
{}

\author{Stephen Kobourov}
{Technical University of Munich}
{stephen.kobourov@tum.de}
{0000-0002-0477-2724}
{}

\author{Jonah Lotz}
{University of Arizona}
{jlotz@arizona.edu}
{}
{}

\author{Jacob Miller}
{Technical University of Munich}
{jacob.miller@tum.de}
{}
{}

\author{Hamlet Taraz}
{University of Arizona}
{hamlet@arizona.edu}
{}
{}

\Copyright{Reyan Ahmed, Cesim Erten, Stephen Kobourov, Jonah Lotz, Jacob Miller, Hamlet Taraz}



\nolinenumbers 

\begin{document}







\maketitle
\thispagestyle{plain}
\pagestyle{plain}

\begin{abstract}
The \normalizedstress metric is widely used to assess graph drawing quality, measuring how closely distances between vertices in a layout match their graph-theoretic distances. This metric is a standard for both evaluation and optimization in many popular graph layout algorithms. However, \normalizedstress is highly sensitive to the scale (size) of the drawing, leading to potentially misleading comparisons between layouts produced by different algorithms. Uniformly scaling a layout can significantly alter stress values without changing the underlying structure, even to the extent that a clearly superior layout can appear to have a higher stress score than a random layout. Although this issue is recognized within the network visualization community, it is rarely addressed with sufficient detail in publications, resulting in critical calculation errors in recent studies.
In this paper, we systematically examine various stress metrics used in the literature and demonstrate that commonly used metrics are affected by layout scale, undermining their reliability for comparison. We identify scale-invariant alternatives  and propose \minoptstress for fair stress-based evaluation.
\end{abstract}

\begin{figure}[ht]
  \centering
    \includegraphics[width=0.95\linewidth]{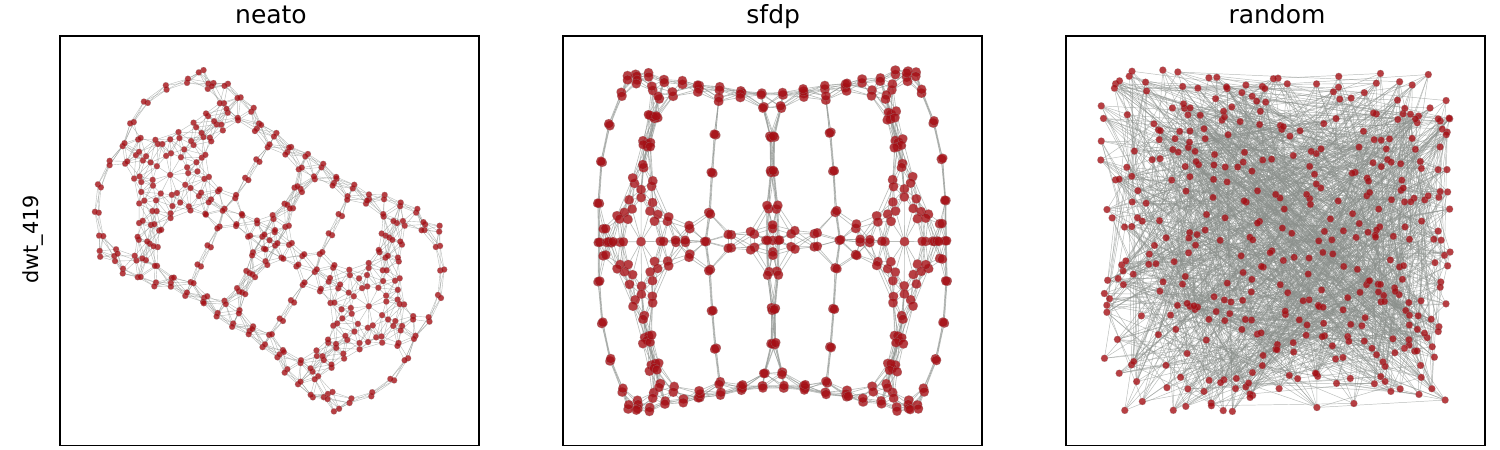}

    \includegraphics[width=0.49\linewidth]{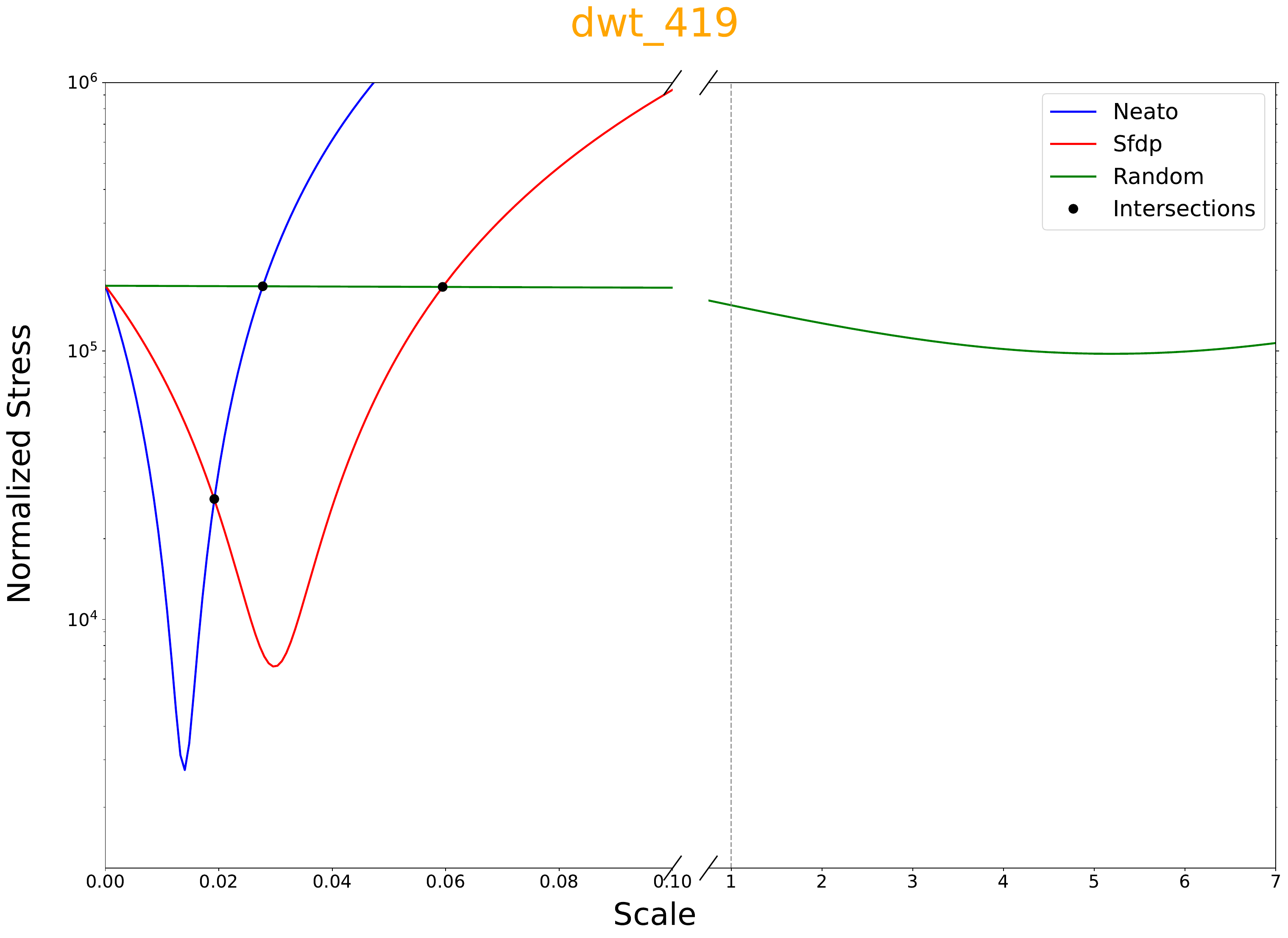}
    
    \caption{
    The dwt\_419 graph drawn by three different algorithms (NEATO~\cite{gansner2005graph}, SFDP~\cite{ellson2001graphviz}, random; left-to-right). At the default scale (1 on the x-axis), the drawing obtained by the random algorithm appears to have the lowest \normalizedstress despite clear visual evidence to the contrary. In fact, by varying the scale of the drawings, any possible ordering of the three algorithms with respect to \normalizedstress values can be found.
    }
    \label{fig:combined-teasers}
\end{figure}

\section{Introduction}
When computing a drawing of a graph, the positions of vertices are typically represented by Cartesian coordinates with respect to some abstract axes. An undirected graph $G = (V,E)$ is a mathematical object made up of a set of objects called vertices, $V$, and a set of connections between them called edges, $E \subseteq V \times V$. The graph layout (embedding) problem involves assigning positions to the vertices, typically in the 2D Euclidean space, subject to predefined optimization goals, such as maintaining graph-theoretic distances or showing clusters~\cite{tamassia2013handbook}. We restrict ourselves to two-dimensional, straight-line layouts, and refer to them as \textbf{\textit{graph drawings}}. A large corpus of algorithms to produce such drawings exists; see~\cite{di2024evaluating} for a recent survey. 

There exists a large and diverse set of graph drawing algorithms, but they all differ in their treatment of coordinate units - some algorithms aim for unit-length edges~\cite{gansner2005graph}, while others may treat vertices as objects with size and mass in an $n$-body simulation~\cite{bostock2011d3}. This difference in treatment results in different numerical sizes of their internal representations. It is intuitively understood by visualization designers that whatever units the output of a layout algorithm is, one must ``scale'' (stretch or shrink) the coordinates of the vertices uniformly, so that it visually fits the desired computer window or canvas. However, when applying evaluation metrics to compare two or more algorithms, this difference in units may have a drastic, out-sized effect on the results if the metric compares distances. In the literature, this effect is typically glossed over in a few sentences~\cite{gansner2012maxent,zhong2023force} if it is accounted for at all.

One such evaluation metric affected by scale is known as \textit{stress}, which is a measure of how closely the embedded geometric distances align with the graph-theoretic distances between vertices. A graph drawing can be represented by a matrix produced from a layout algorithm, $X \in \mathbb{R}^{|V| \times 2}$, where row $X_i$ is the position of vertex $v_i$ in the drawing. Stress-based graph layout algorithms also use the all-pairs-shortest-path matrix $D \in \mathbb{R}^{|V| \times |V|}$, where $d_{ij}$ is the length of the shortest path between $v_i$ and $v_j$. When evaluating the quality of a layout, we often compare the graph-theoretic distances in $D$ with the Euclidean distances implied by $X$. This comparison is captured by a family of \textit{stress metrics}, defined formally in \autoref{sec:stress-defs}.
Stress is the third most frequently used layout evaluation metric (after runtime and crossings) in the literature~\cite{di2024evaluating}. While there are many stress variants, the most frequently used is the so-called  \textit{\NormalizedStress} ($\Normalizedstressmath$)~\cite{brandes2008experimental, gansner2005graph, gansner2012maxent,ortmann2016sparse,kruiger2017graph}. As typically written, \normalizedstress is severely affected by scale:
\begin{align}
\label{eq:loss-stress}
\Normalizedstressmath(X,D) = \sum\limits_{i<j}\;d_{ij}^{-2}(||X_i - X_j||_2 - d_{ij})^2 
\end{align}
The normalization factor $d_{ij}^{-2}$ balances the influence of short and long distances: the greater the graph-theoretic distance, the more tolerance we allow in the difference between the two distances. However, despite the name and presence of this normalization factor, the $\Normalizedstressmath$ metric is not normalized with respect to the scale of the positions of vertices, $X$, nor is it normalized to a range (which is unbounded). Fixing $X$ and $D$ as constants, we can observe the behavior of the function $\Normalizedstressmath(\alpha X, D)$ when varying the scaling factor $\alpha > 0$ in \autoref{fig:combined-teasers}. Scaling the embedding (e.g., by $\alpha = 1, \alpha=2, $ or $\alpha=0.5$) does not change the geometric relationship between the vertices at all. However, as we can see from \autoref{fig:combined-teasers}, such scaling affects the value of $\Normalizedstressmath$ drastically. This makes $\Normalizedstressmath$ a poor measure to compare any two layout algorithms whose outputs might differ in scale. 

Further, when scale is not accounted for, one cannot say where on the $\Normalizedstressmath$-scale plot a drawing might fall. In \autoref{fig:combined-teasers}, this amounts to picking \textit{\textbf{any}} point on the blue line for NEATO, \textbf{\textit{any}} point on the red line for SFDP, and again \textbf{\textit{any}} point on the green line for the random algorithm. By doing this, any desired order of the drawing algorithms could be obtained, either by accident or by intentionally choosing desirable scale factors. This sensitivity to scale has been observed in the literature~\cite{gansner2012maxent,wang2023smartgd,Wageningen2024,zhong2023force}, but is almost never explicitly or clearly addressed.

While the effect of scale on stress is known to some experts in the field, it is not well described or clearly documented in the literature;
 see Table~\ref{tab:survey}.
Scale-invariant metrics have recently been considered in the context of dimensionality reduction methods~\cite{smelser2024}, but no such work covers the specific concerns of the graph drawing community.
In the graph drawing domain, a few papers mention attempts to ``resize the layout to achieve minimum stress''~\cite{gansner2012maxent,kruiger2017graph}. Other papers define the scale-sensitive \normalizedstress of~\autoref{eq:loss-stress}, but actually implement \minoptstress in their publicly available code~\cite{DBLP:conf/gd/MillerHK23,xue2022taurus}.
This lack of specificity has led to incorrect implementations in libraries~\cite{nollenburg_et_al:LIPIcs.GD.2024.45, jeon23vis} and paper evaluations (\autoref{sec:experiment1}). Personal communications with practitioners and experts also confirm a gap in understanding of scale sensitivity, often describing ``normalization'' that does not accurately account for scale.

We believe the majority of graph layout researchers understand and account for scale when computing stress. However, the above factors create a poor environment surrounding quantitative evaluation of graph layouts. It makes it unnecessarily difficult for newcomers to intuit what was actually meant when the stress metric is described, inviting errors. It harms reproducibility whether or not the code is provided, as even open source code is likely to succumb to link-rot given enough time. It is also simply bad scientific practice to be vague when specifying a computation. 

Our goal  is to provide an overview of the use of the stress metric in graph drawing, thoroughly explain why it is necessary to account for scale, demonstrate that not accounting for scale has affected evaluation results, and show experimentally that scale-invariant metrics should always be preferred.

\section{Survey of Stress as an Evaluation Metric}
\label{sec:survey}

%

The notion of stress originates from the field of statistical analysis, where it is used as a dimension reduction (DR) technique. The class of DR algorithms that directly optimize stress is known as Multi-Dimensional Scaling (MDS). 

{There are three primary variants of multidimensional scaling (MDS) that have been developed and applied to various tasks. \textbf{Classical MDS}~\cite{torgerson1952multidimensional}, the earliest method, assumes that distances are derived from a Euclidean space, which enables a closed-form solution. It is often employed as an initialization step for more complex graph drawing techniques~\cite{brandes2006eigensolver,koren2005drawing}. \textbf{Metric MDS}~\cite{sammon1969nonlinear} generalizes this approach by defining stress as a differentiable objective function, which can be minimized using gradient-based optimization; this makes it particularly useful in graph drawing applications~\cite{kamada1989algorithm,gansner2005graph,zheng2018graph}. \textbf{Non-metric MDS}~\cite{kruskal1964multidimensional}, in contrast, focuses on preserving the rank order of distances rather than the actual values. However, it has not seen widespread adoption in graph layout contexts.
}

The popularity of stress is in part due to the fact that it is a simple to explain \textit{faithfulness} metric~\cite{nguyen2013faithfulness,eades2015shape}. Faithfulness metrics attempt to capture how well a graph drawing represents structural properties of the underlying graph. 
Prior work has shown that not only do readers prefer drawings with lower stress~\cite{chimani2014people}, but they can also reliably identify drawings with lower stress~\cite{mooney2024perception} making it a desirable quality of a graph drawing. 
The family of faithfulness metrics includes neighborhood preservation~\cite{kruiger2017graph} and shape-based metrics~\cite{eades2015shape}, but the most frequently used and evaluated is stress.


Recent advances in graph drawing have increasingly incorporated machine learning (ML) techniques to improve layout quality, automate parameter tuning, and adapt layouts to diverse tasks and data types. For instance, frameworks such as SPX~\cite{devkota2019stress}, (GD)$^2$~\cite{ahmed2020graph}, (SGD)$^2$~\cite{ahmed2022multicriteria}, SmartGD~\cite{wang2023smartgd}, CoRE-GD~\cite{grotschla2024core}, and others leverage ML to learn effective graph embeddings that better capture structural properties and user preferences. These approaches typically optimize layout quality metrics, with stress or normalized stress variants often serving as core evaluation criteria.
Without addressing scale sensitivity, comparing and benchmarking different approaches using normalized stress becomes unreliable.
To provide an overview of stress-based evaluations, we survey recent network visualization papers.

\subsection{Survey Methodology}
We identified a ``root" paper, that is highly cited and among the first to employ stress as an evaluation metric: the paper on stress majorization by Gansner et al.~\cite{gansner2005graph}, as it was the typical reference for \normalizedstress found in the literature. Then, we used the OpenCitation API~\cite{10.1162/qss_a_00023}  to perform a breadth first search on the citation network of the paper, to a depth of four. This resulted in over 4,000 papers, which were then filtered by keywords, excluding papers whose titles or abstracts did not include one of the following substrings: ``graph'', ``network'', ``mds'', ``layout'', ``embed'', ``drawing'', as these were terms found in relevant papers we were familiar with. This resulted in around 1200 papers. From these, we selected papers from the last 5 years (about 400), plus some other notable publications (identified by having been cited in reference to stress). We further filtered out papers that do not discuss or evaluate stress which resulted in 44 papers. We provide an annotated list of these papers in our repository. We categorize each paper into the following categories. 
\begin{itemize}
    \item \textbf{Explicit Scale-normalization} papers have an explicit definition of how they account for scale when measuring stress. This is the ideal case, it is clear, precise, and reproducible given the content of the paper.
    \item \textbf{Brief Scale-normalization} papers have a short description of how scale is accounted for, e.g. ``the layout is scaled to achieve the minimum stress.'' Notably, the written mathematical formula does not reflect what is being computed, which leaves too much room for interpretation.
    \item \textbf{Empirical Scale-normalization} papers are those which define \normalizedstress as in \autoref{eq:loss-stress}, but implement some type of scale normalization (verified via publicly available code, or code provided by the authors). 
    \item \textbf{Unknown Scale-normalization} papers define some variant of~\autoref{eq:loss-stress} with no mention of accounting for scale, and we were unable to verify the implementation.
\end{itemize}

\begin{table*}
\centering
    \bgroup
    \def\arraystretch{2}
    \begin{tabular}{|c c|}
        \hline 
        Explicit Scale-normalization &  \cite{Heiter_2022,wang2023smartgd,Wageningen2024,Simonetto2018,Simonetto_2020,arleo2022} \\ \hline
        Brief Scale-normalization &  \cite{Brandes_2019, Das_2019,Rahman_2020,zhu2020drgraph,gansner2012maxent,zhong2023force,ortmann2016sparse,welch2017measuring} \\ \hline
        Empirical Scale-normalization & \cite{kruiger2017graph,DBLP:conf/gd/MillerHK23, xue2022taurus,ahmed2020graph,ahmed2022multicriteria}\\ \hline
        Unknown Scale-normalization & \parbox[c]{0.3\textwidth}{\cite{Marc_lio_Jr_2019,devkota2019stress,zheng2018graph,Coimbra_2021,Boyarski_2021,Dzemyda_2021,espadoto2019toward,Sheng_2021,Olauson_2020,Park_2020,Teixeira_2020,Yu_2022,Xiao_2022,Miller_2023,Giovannangeli_2024,Sheng_2019,Cai_2021,Giovannangeli_2021,Meidiana_2020,DBLP:journals/cga/WangYHS21,Gray_2022,Hong2019,wang2017revisiting,DBLP:conf/gd/KobourovPS14}} \\ \hline
    \end{tabular}
    \egroup
    \vspace{0.25cm}
    \caption{Literature survey results, showing a large number of papers with unclear treatment of scale.}
    \label{tab:survey}
\end{table*}

\subsection{Survey results}
We include an analysis of our survey results in this section, summarized in \autoref{tab:survey}.

\subsubsection{Explicit Scale-normalization}
These are papers that use a scale-aware stress implementation and clearly explain this in the write-up. The first two appeared within a few weeks of each other in 2022 and both give the formula for the minimum scaling factor~\cite{Heiter_2022,wang2023smartgd}. Wageningen et al. followed sometime later with the same formula~\cite{Wageningen2024}. It is worth mentioning that this formula (\autoref{eq:sns-alpha}) was known in the community well before it first appeared in the literature in 2022 (as early as the 1990s, as confirmed by personal communication with several experts). 

\subsubsection{Brief Scale-normalization}
This category is fairly common in the literature. A typical paper in this group mentions in passing that the stress function is sensitive to scale and that the ``minimum stress" should be selected, while still providing the definition of \normalizedstress in~\autoref{eq:loss-stress}. Papers in this category leave too much room for interpretation and might help reinforce the incorrect belief that ``normalized stress'' is scale-invariant.

Many papers in this category cite the 2012 paper by Gansner et al.~\cite{gansner2012maxent}, which to our knowledge is the first paper to explicitly state and address the issue of scale. However, this paper also only briefly mentions the issue of scale and does not show the actual solution: neither how to compute the appropriate scale, nor how to modify the \normalizedstress function so it becomes scale-invariant. 

\subsubsection{Empirical Scale-normalization}
A paper in this category does not mention or discuss the issue of scale, while implementations associated with the paper do attempt to take scale into account. Several implementations we found were rather inefficient. Some used a binary search~\cite{DBLP:conf/gd/MillerHK23}, while others chose a minimum over a large spectrum of sizes~\cite{Simonetto2018,Simonetto_2020,arleo2022}.

We have also seen three ways of dividing the resulting stress value: no division at all, $|V|$, and $\binom{|V|}{2}$. Dividing \autoref{eq:loss-stress} by these values gives you the total stress, average stress per node, and average stress per node pair respectively. While this does not affect the evaluation per se (as long as the value is consistent, it will not affect the ranking), this affects the range of values the reported stress will take. Ideally, the reported function detailed in the paper is \textit{exactly} what is computed.

Papers of this type can be confusing, as the description in writing does not match the actual implementation. Trying to replicate results in such papers becomes difficult, if not impossible. 

\subsubsection{Unknown Scale-normalization}
Papers in this category are the most confusing with respect to stress computations. Stress is defined or used without mention of scale, leaving the treatment of drawing sizes completely ambiguous. This category includes papers that use scale as an evaluation metric but do not fall into the above categories. Some papers in this category assume familiarity with stress, providing it without reference as a ``well-known" aesthetic metric~\cite{Giovannangeli_2021} or citing one of either~\cite{gansner2005graph,kamada1989algorithm} where stress was first defined in the graph layout literature, as in~\cite{ahmed2022multicriteria}. Other papers define it as an optimization function (where the scale is more or less irrelevant), but go on to later use it as an evaluation metric~\cite{devkota2019stress}. Others use ``scaled" when referring to stress, but it is unclear whether this refers to geometric scale/size, the average, or something else~\cite{eades2015shape}.

\subsection{Survey Discussion}
The key takeaway from this survey is that an important quality and faithfulness metric is not applied correctly or consistently.
Of the 44 surveyed papers, only six had a clear treatment of scale. Eight papers had some discussion of scale but had an overall ambiguous write-up. Five papers addressed scaling issues in their implementation but did not mention this at all in the write-up. The remaining 25 of 44 papers did not discuss scale in their evaluation or definitions. 

Reproducibility, in a scientific context, means that a study's results can be recreated by another researcher using the described methods and data. Our survey suggests that only 15\% of the 44 papers pass this test. 
With this in mind, we aim to clarify how stress is affected by scale, to what degree this can affect evaluation results, and what scale-invariant measures should be used. We also provide an efficient implementation of scale-normalized stress which will hopefully replace the many scale-sensitive variants currently in use.

\section{Stress-based Quality Metrics}
\label{sec:stress-defs}

\begin{table*}[ht]
\begin{center}
\begin{tabular}{|c  c  c l|}
\hline
Stress Metric & Definition & \parbox[c]{1.5cm}{\centering \vspace{2pt}Invariant\newline to scale?\vspace{2pt}} &  Range \\
\hline

 \textbf{\Rawstressmath} & $\sum\limits_{i<j}\;(||X_i - X_j|| - d_{ij})^2$ & No &  [0, $\infty$)\\  
 \hline 
 \textbf{\Normalizedstressmath} & $\sum\limits_{i<j}\;d^{-2}_{ij}(||X_i - X_j|| - d_{ij})^2$ & No & [0, $\infty$) \\ 
 \hline\hline
 \textbf{\Minoptstressmath} & $\sum\limits_{i<j}\;d^{-2}_{ij}(\min_{\alpha > 0} \alpha||X_i - X_j|| - d_{ij})^2$ & Yes &  [0, $\infty$)  \\ 
 \hline
 \textbf{\Shepardgoodnessmath} & Spearman rank correlation $||X_i - X_j||$, $d_{ij}$ & Yes &  [-1,1] \\ 
 \hline
 \textbf{\Kruskalstressmath} & $\sqrt{\frac{\sum\limits_{i<j}\;(||X_i - X_j|| - \hat{d}_{ij})^2}{\sum\limits_{i<j}\;||X_i - X_j||}}$ & Yes &  [0,1]\\ 
 \hline 
\end{tabular}
\end{center}
   \caption{Overview of the stress metrics under consideration. The acronyms used in the leftmost column refer to those introduced in \autoref{sec:scale-sensitive-stress-defs} and \autoref{sec:scale-invariant-stress-defs}.
}
   \label{tab:stress_table}
\end{table*}



In this section, we describe several stress-based quality metrics identified from our survey in~\autoref{sec:survey}. We provide the definitions, and show both analytically and empirically whether or not the measured value changes with respect to scaling of a graph drawing. For the remainder of this section, we define a stress metric as a two-parameter function $M(X,D)$ where $X \in \mathbb{R}^{|V| \times 2}$ is a drawing and $D \in \mathbb{R}^{|V| \times |V|}$ is a graph distance matrix. We can then observe the behavior of $M(\alpha X, D)$ as we vary $\alpha > 0$. A scale-sensitive metric is any metric that changes when $\alpha$ changes, while a scale-invariant metric is one that does not change with $\alpha$.

\subsection{Scale-sensitive Stress Metrics}
\label{sec:scale-sensitive-stress-defs}

\begin{figure}[t]
    \centering
    \begin{minipage}{0.98\textwidth}
        \centering
        \includegraphics[width=\linewidth]{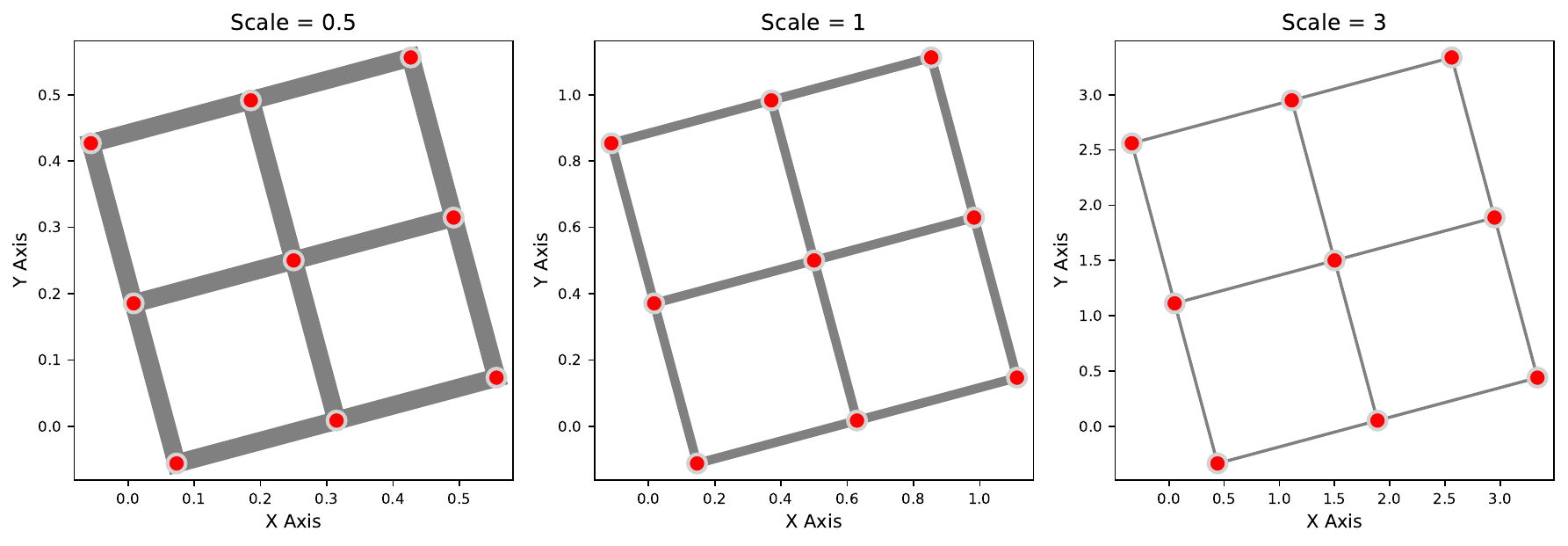}
        \subcaption{\centering}
        \label{fig:scale-illust}
    \end{minipage}%
    \hfill
    \begin{minipage}{0.98\textwidth}
        \centering
        \includegraphics[width=\linewidth]{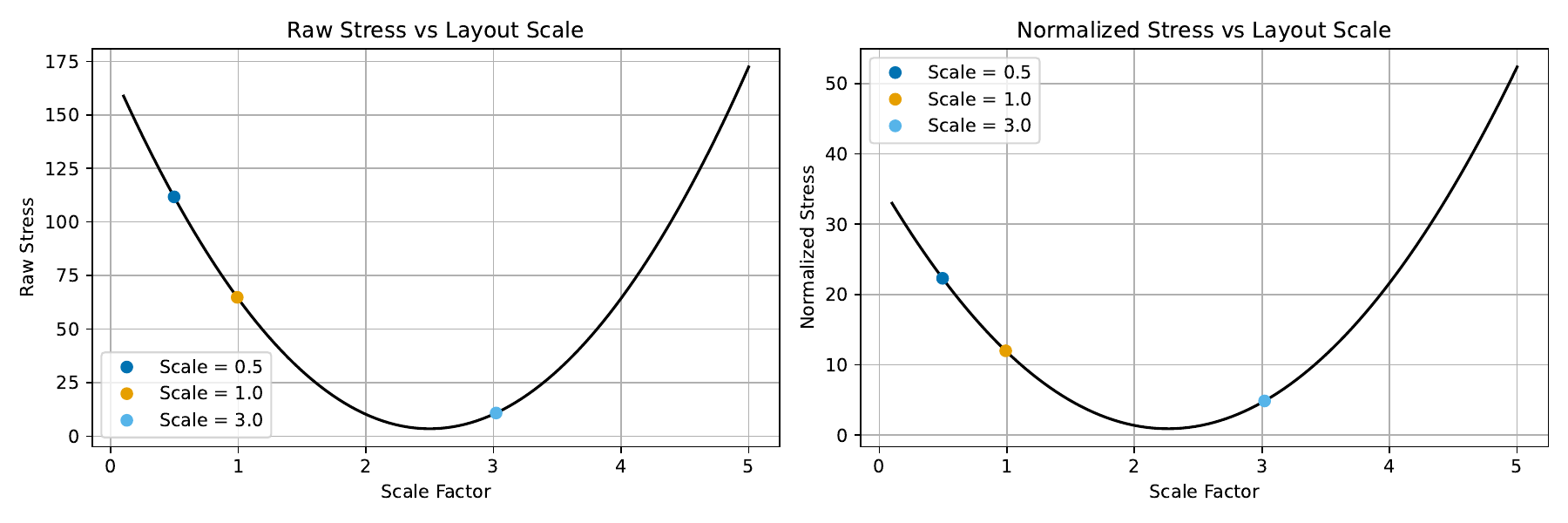}
        \subcaption{\centering}
        \label{fig:scale-changing}
    \end{minipage}
    \caption{
    (a) Scaling the size of a drawing does not affect its structure or connectivity, as shown with the same rotated grid graph at half size (left), default size (middle), and three times the size (right). Pen width is unadjusted for clarity.
    (b) Plots of scale-sensitive stress metrics as the drawing size changes. The curves correspond to the small grid graph in \autoref{fig:scale-illust}. Raw stress decreases with scale (0.5: 111.28, 1.0: 64.07, 3.0: 10.23), normalized stress also decreases (0.5: 22.19, 1.0: 11.82, 3.0: 4.64).
    }
    \label{fig:combined}
\end{figure}

\textbf{\RawStress(\Rawstressmath):}
\Rawstress (also known as unweighted stress) as defined by Sammon~\cite{sammon1969nonlinear} contains no extra terms or normalizing factors, but is simply the sum of squared differences of ideal and realized differences.
It has been used in various studies~\cite{brandes2008experimental,buja2008data}. However, a large problem with \rawstress is that it over-penalizes large distances in the drawing and under-penalizes short distances~\cite{gansner2012maxent}.
It is defined formally as:
\begin{equation}
    \label{eq:raw-stress}
    \Rawstressmath(X,D) = \sum_{i<j}(||X_i - X_j || - d_{ij})^2
\end{equation}
\Rawstress is clearly a scale-sensitive metric as can be seen in \autoref{fig:scale-changing} (left).
In fact, we can rewrite it as a quadratic polynomial with respect to $\alpha$:
\begin{align*}
\Rawstressmath(\alpha X, D) &= \sum_{i<j}(\alpha||X_i - X_j || - d_{ij})^2 \\
&= \alpha^2\sum_{i<j}||X_i - X_j||^2 - 2\alpha\sum_{i<j}||X_i - X_j||d_{ij} + \sum_{i<j}d_{ij}^2
\end{align*}

\medskip\noindent
\textbf{\NormalizedStress (\Normalizedstressmath):}
The definition of \normalizedstress ($\Normalizedstressmath$) is provided in \autoref{eq:loss-stress}.
This metric comes from a desire to ``normalize'' the penalty of short and long distances by the factor $w_{ij}$ which is usually set to $d_{ij}^{-2}$~\cite{gansner2005graph}. This ensures distance is weighted analogously to an inverse-square law. Despite being the most popular formulation of stress to evaluate graph drawings, $\Normalizedstressmath$ makes for a poor metric to compare the output of two different algorithms without accounting for scale. This holds true even for very small graphs, such as the grid graph shown in \autoref{fig:scale-illust}. Of course, when used as an optimization function, \normalizedstress produces nice drawings, as has been shown in many papers e.g.~\cite{gansner2005graph, zheng2018graph}.

However, when used as a quality metric, one can achieve nearly any order desired when comparing drawing algorithms depending on the scale they are evaluated with. A bad actor could quietly change the scale of drawings to achieve a desirable outcome for an algorithm. \Normalizedstress, as shown in \autoref{fig:scale-changing} (right)
is scale-sensitive. We observe it as a function of $\alpha$, for convenience later: 
\begin{equation}
    \label{eq:ns-curve}
    \Normalizedstressmath(\alpha X, D) = \sum_{i<j} \; d_{ij}^{-2}(\alpha ||X_i - X_j|| - d_{ij})^2
\end{equation}

Clearly, the value of $\Normalizedstressmath$ changes with respect to $\alpha$, and it can be rewritten as a quadratic polynomial of $\alpha$ using a similar derivation as \autoref{eq:raw-stress}. It is also easy to see that the so-called ``normalization" term of $d_{i,j}^{-2}$ does not account for the size of the drawing.

Using this property, one can manipulate the order of the quality of different layouts as described in \autoref{fig:combined-teasers}. 
We study the behavior of \autoref{eq:ns-curve} more carefully, to show just how much \normalizedstress varies with scale. Two drawings of the same graph, $X$ and $X'$ define two quadratic curves in $\alpha$, so they can intersect (and thus, the order of evaluation metrics can change) at most twice. 
One of these intersections is trivially at $\alpha = 0$, and the scaling factor must strictly be positive. The other intersection can be computed as follows.
Let $\Normalizedstressmath(\alpha X,D)$ and $\Normalizedstressmath(\alpha X',D)$ be the quadratic $\Normalizedstressmath$ functions computed for two fixed layouts $X$ and $X'$, respectively. The intersection occurs where $\Normalizedstressmath(\alpha X,D)=\Normalizedstressmath(\alpha X',D)$, so by solving for $\alpha$ one can find where the order changes:
\begin{equation}
    \label{eq:intersect-alpha-norm}
    \alphaintnorm = 2\frac{\sum\limits_{i<j}\;d_{ij}^{-1}(||X_{i}-X_{j}||-||X'_{i}-X'_{j}||)}
    {\sum\limits_{i<j}\;d_{ij}^{-2}(||X_{i}-X_{j}||^2-||X'_{i}-X'_{j}||^2)}
\end{equation}

Furthermore, for $k$ drawings of the same graph, each drawing defines a quadratic equation of $\Normalizedstressmath$ in terms of $\alpha$. Hence, $2\binom{k}{2}$ intersections are generally possible; however, half of them occur at $\alpha=0$. The remaining $\binom{k}{2}$ intersections are all non-zero. Still, at most $\binom{k}{2}$ effective intersections are possible. This means that when comparing the embeddings from $k$ algorithms for a given graph, there are $\binom{k}{2}+1$ different orderings of \normalizedstress. This  is surely undesirable for any quality metric, as one can easily change the order of layouts by modifying the scale. One could also further manipulate the orderings of layouts by scaling each layout differently. For example, in  \autoref{fig:combined-teasers}, a random graph drawing scaled beyond 0.06 has a better \normalizedstress score than the stress-optimized layouts.




\subsection{Scale-invariant Stress Metrics}
\label{sec:scale-invariant-stress-defs}

\textbf{\MinoptStress (\Minoptstressmath):} 
In order to compare the quality of layouts obtained by different algorithms we should use the minimum stress score over all sizes of drawings of \normalizedstress. 
Since \normalizedstress corrects raw stress for unbalanced graph distances and finding the minimum stress value across all sizes corrects for unbalanced embedding distances, we refer to this metric as  {\bf \MinoptStress}.

\MinoptStress has been implicitly or explicitly mentioned or used in only a handful of papers~\cite{welch2017measuring, gansner2012maxent, zhong2023force,zhu2020drgraph}. With very few (also very recent) notable exceptions~\cite{wang2023smartgd, Wageningen2024}, the derivation of \MinoptStress has not been spelled out. With this in mind, here we define explicitly how this should be done. The scalar $\alpha_{min}$ which finds the size of drawing that minimizes \normalizedstress can be directly computed as follows:



\begin{equation}
    \label{eq:sns-alpha}
    \alpha_{min} = \frac{\sum\limits_{i<j}\;d^{-1}_{ij}||X_i-X_j||}{\sum\limits_{i<j}\;d_{ij}^{-2}||X_i-X_j||^2}
\end{equation}

This leads to a succinct formal definition of \minoptstress:

\begin{equation}
\label{eq:NS_min}
\Minoptstressmath = \Normalizedstressmath(\alpha_{min} X, D) = \sum_{i<j} \; d_{ij}^{-2}(\alpha_{min} ||X_i - X_j|| - d_{ij})^2
\end{equation}

\begin{figure}[!htbp]
    \centering
    \begin{minipage}{0.99\linewidth}
        \centering
        \includegraphics[width=\linewidth]{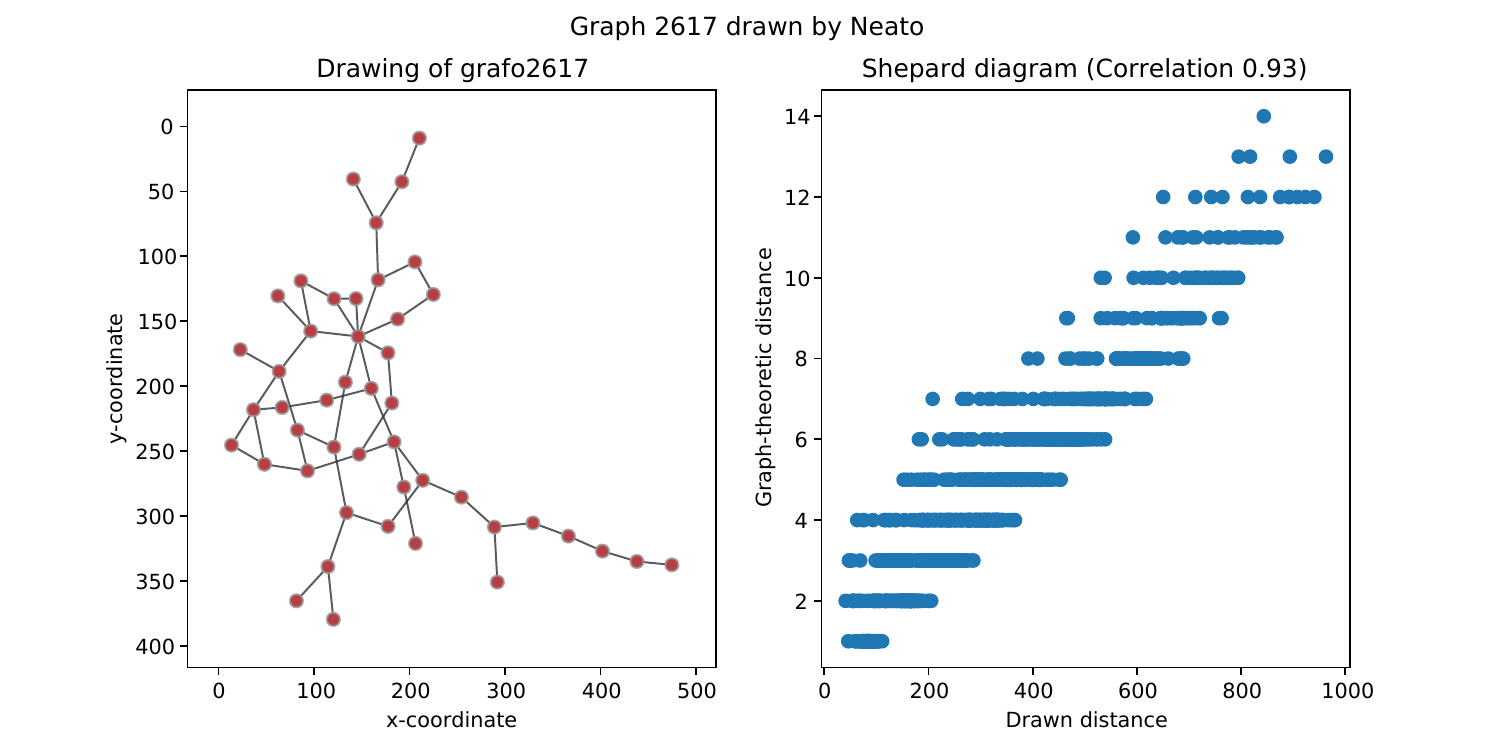}
    \end{minipage}%
    \\
    \begin{minipage}{0.99\linewidth}
        \centering
        \includegraphics[width=\linewidth]{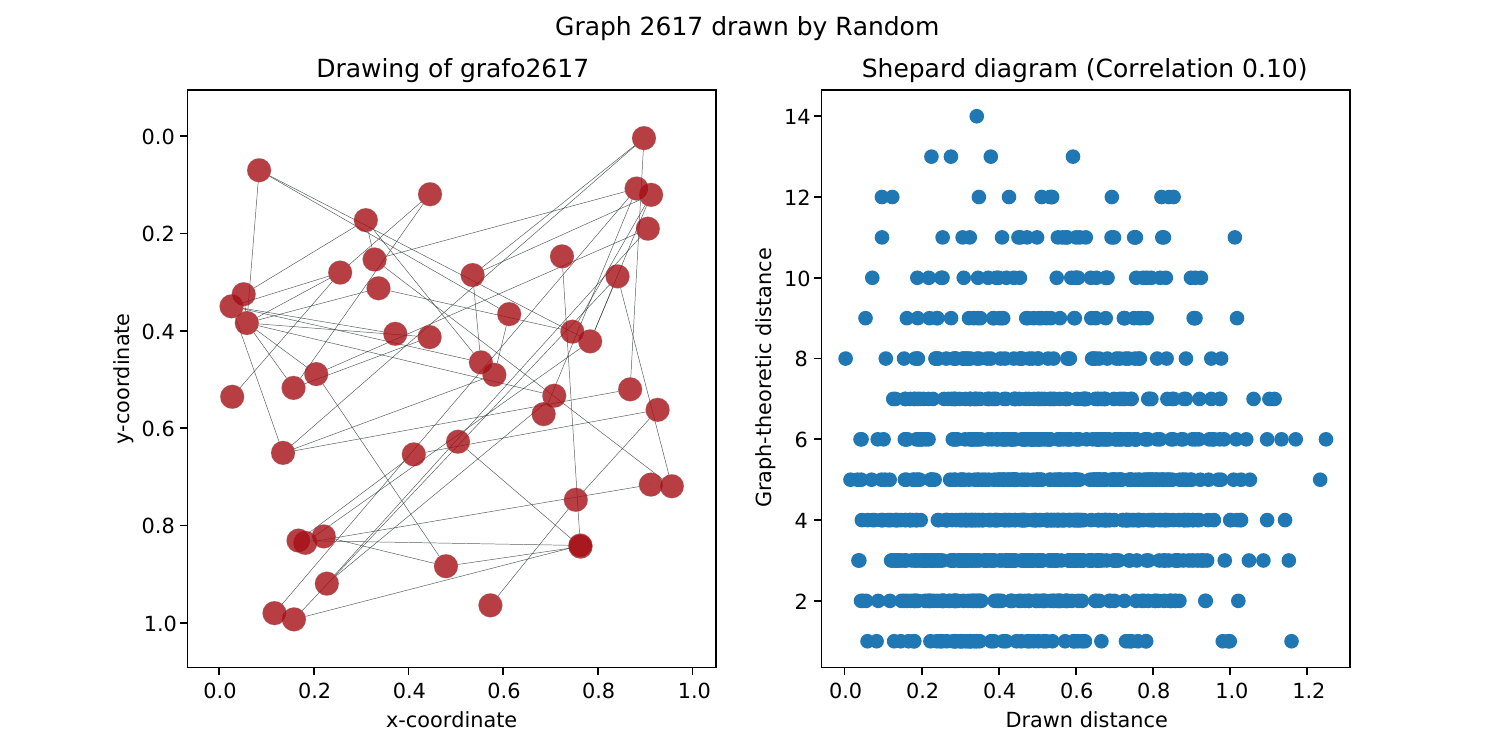}
    \end{minipage}
    \caption{Two drawings and the accompanying Shepard diagrams for graph grafo2617. The left, stress-based drawing by NEATO places more points of the Shepard diagram closer to the diagonal than the Random drawing. This is reflected in the correlation coefficients: 0.9 for NEATO and 0.1 for Random.}
    \label{fig:shepard-illust}
\end{figure}

\medskip\noindent
\textbf{\ShepardGoodness (\Shepardgoodnessmath)}: The Shepard diagram~\cite{shepard1962analysis} was introduced in 1962 as a measure of stress when performing dimensionality reduction. It is a scatter plot that compares pairwise distances in the input high dimensional space with the pairwise distances in the output low dimensional space. It can be naturally adapted to graph embeddings by using the pairwise graph distances as the input and the Euclidean distances in the layout as the output. Specifically, each point in the scatter plot $c_{ij}$ corresponds to a pair of vertices, with the location of each point encoding the graph-theoretic distance in one coordinate and the drawing distance between the pair in the other coordinate: $c_{ij} = (||X_i - X_j||, d_{ij})$.

If all $|V|^2$ points are collinear and the line has slope $45^{\circ}$ we have a  perfect drawing, in which every graph-theoretic distance is exactly the drawn distance. While this is not possible in general, configurations of the diagram which are closer to a straight line are more faithful than those which are more uniformly distributed. In fact, \normalizedstress and \rawstress can be computed directly from this diagram, by measuring the distance in $x$-coordinate from each point to the line $y=x$. Example diagrams are shown in \autoref{fig:shepard-illust}; note that the better layout has a Shepard diagram closer to the diagonal.

The primary assessment for the quality of such a diagram is the computation of its Spearman rank correlation between its $x$ and $y$ components. This correlation, termed the “Shepard Goodness of Fit”, quantifies the monotonic relationship between the graph and embedding distances. The value of this correlation ranges between -1 and 1, with these two endpoints representing a perfect positive and negative correlation of distances, respectively. This value is inherently scale-invariant as it relies solely on the rank ordering of distances rather than their absolute magnitudes. This characteristic ensures that the metric remains unaffected by linear transformations of the data, facilitating consistent evaluations across scaling factors.

\medskip\noindent
\textbf{\KruskalStress (\Kruskalstressmath)}:
The \kruskalstress introduced by Kruskal~\cite{kruskal1964multidimensional} is not concerned with strictly preserving exact distances in an embedding, but rather the ranking (ordering) of the distances. For example, for any given vertex, the closest, second closest, and third closest vertices should be the closest, second closest, and third closest in the drawing (arbitrarily broken for ties). This metric has been studied in the dimensionality reduction literature~\cite{espadoto2019toward}, but has not, to our knowledge, been applied as an evaluation metric for graph layouts. 

\begin{figure}[!htbp]
    \centering
    \begin{minipage}{0.99\linewidth}
        \centering
        \includegraphics[width=\linewidth]{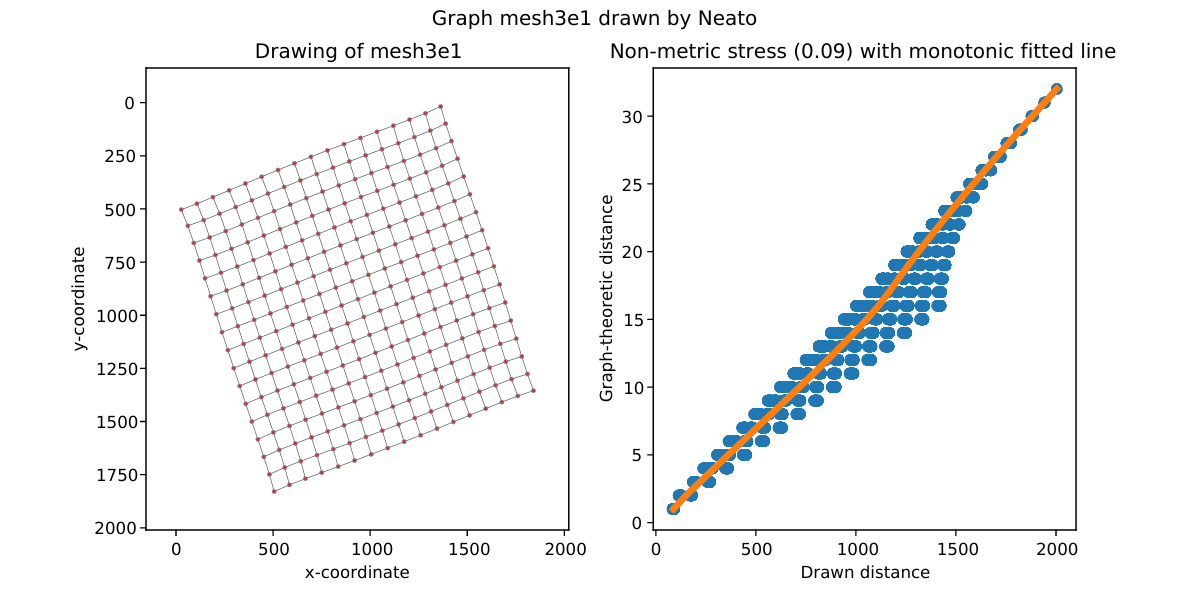}
    \end{minipage}%
    \\
    \begin{minipage}{0.99\linewidth}
        \centering
        \includegraphics[width=\linewidth]{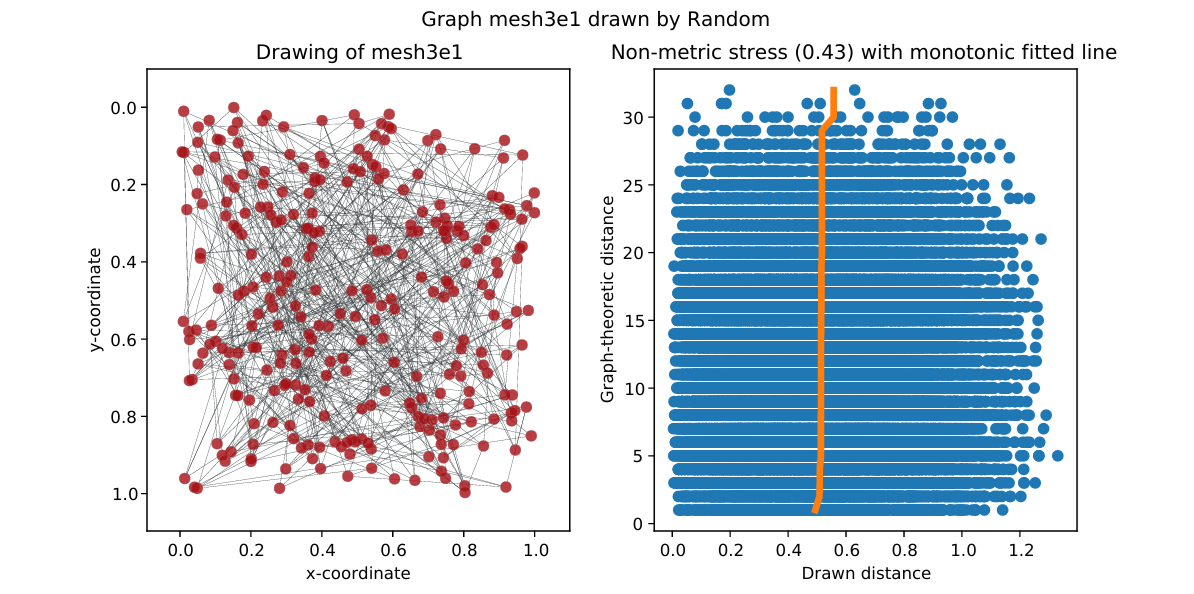}
    \end{minipage}
    \caption{Illustration of the first step of computing \kruskalstress. From a drawing (left), a Shepard diagram is computed (right) and a monotonic regression is performed on its components (the orange line). Configurations in which the average horizontal distance from the blue points to the orange line is small have lower \kruskalstress.}
    \label{fig:kruskal-illust}
\end{figure}

The formulation of non-metric stress appears deceptively similar to metric stress, but with several key differences. 
Firstly, we need to compute the matrix $\hat{D}$ from $X$ and $D$ as follows:
\begin{compactenum}
    \item Compute the Shepard diagram to obtain coordinates for each pair of vertices: $c_{ij} = (||X_i - X_j||, d_{ij})$.
    \item Compute the monotonically increasing line of best fit (isotonic regression) with $\binom{|V|}{2}$ degrees of freedom.
    \item Fill $\hat{D}$ with the distances $\hat{d}_{ij}$ in the $x$-coordinate from the fitted line to $c_{ij}$.
\end{compactenum}
Then, \kruskalstress is computed as:
\begin{align}
    \Kruskalstressmath(X,D) = \sqrt{\frac{\sum\limits_{i<j}\;(||X_i - X_j|| - \hat{d}_{ij})^2}{\sum_{i<j}||X_i - X_j||^{2}}}
    \label{eq:Kruskal-stress}
\end{align}
Note that \autoref{eq:Kruskal-stress} not only computes the squared differences between the layout distances and the predictions of $\hat{D}$, but also normalizes the squared differences in the numerator by the squared layout distances. A visual example of how the metric is computed is given in \autoref{fig:kruskal-illust}.
The metric is thus scale-invariant, as $\hat{d}_{ij}$ is defined as the distance in the $x$-coordinate from $c_{ij}$ to the fitted line, and the drawing distances make up the $x$-coordinate of the Shepard diagram. When $\alpha$ is applied to $X$, the same factor is also applied to $\hat{D}$: 
\[    
\Kruskalstressmath(\alpha X,D) = \sqrt{\frac{\sum\limits_{i<j}\;(\alpha||X_i - X_j|| - \alpha\hat{d}_{ij})^2}{\sum_{i<j}(\alpha||X_i - X_j||)^{2}}}
\]
\[= \sqrt{\frac{\sum_{i<j}\alpha^2(||X_i - X_j|| - \hat{d}_{ij})^2}{\sum_{i<j}\alpha^2||X_i - X_j||^{2}}}\]


\section{Experimental Evaluation}
Thus far, we have many different definitions of the stress metric, some  scale-sensitive and others scale-invariant. In this section, we detail empirical experiments to emphasize the need to accurately account for scale when measuring stress. 

We first show that not accounting for scale sensitivity can affect  quantitative evaluation results, in a replication experiment. We then show (in a second experiment) that the most frequently used version of stress (\normalizedstress) can give clearly incorrect results on a large dataset of graphs. We further observe the correlation of these metrics, along with their computation time on each drawing. 

\subsection{Replication Experiment}
\label{sec:experiment1}
In previous sections, we have identified conflicting definitions present in the literature which, in theory, can affect evaluation results. We now show that this can indeed also happen in practice.

\subsubsection{Replication Design and Description}
We identified several candidate papers which present a new graph layout algorithm, which as part of their evaluation compute (normalized) stress values on a set of graphs to compare to competing state-of-the-art techniques. It was then necessary to narrow the results down to papers with available implementations that are feasible to verify. We investigate the results of Ahmed et al.~\cite{ahmed2022multicriteria} as they provide both open source code and drawings. The paper itself makes no mention of resizing drawings for stress evaluation, and while a scale-normalized stress implementation exists in the open source code, it is applied inconsistently. We believe this work is representative of typical graph layout technique papers.

This work proposes an algorithm called $(SGD)^2$ that can optimize different drawing criteria, for example, stress (ST), ideal length (IL), neighborhood preservation (NP), crossings (CR), angular resolution (ANR), vertex resolution (VR), etc. The quality of each drawing criterion can be measured using a loss function, and the algorithm can combine multiple loss functions to optimize a combined objective. In one experiment of the paper, eight well-known graphs are optimized on different sets of criteria, and the stress score is reported for each of them. We provide statistics for these eight graphs in Table~\ref{tab:rep_graph_statistics}.

\begin{table}[ht]
    \centering
    \small 
    \begin{tabular}{|l|c c c c|}
    \hline
    & \multicolumn{4}{c|}{\textbf{$(SGD)^2$ Graph Stats}} \\ \hline
    \textbf{Statistic} & \textbf{Mean} & \pmb{$\sigma$} & \textbf{Min} & \textbf{Max} \\ \hline
    $|V|$ & 296.71 & 328.85 & 20.0 & 1005.0 \\ \hline
    $|E|$ & 850.57 & 1258.57 & 30.0 & 3808.0 \\ \hline
    Min Degree & 3.28 & 2.05 & 1.0 & 7.0 \\ \hline
    Max Degree & 9.0 & 8.35 & 3.0 & 28.0 \\ \hline
    Density $|E| / \binom{|V|}{2}$ & 0.045 & 0.048 & 0.0048 & 0.158 \\ \hline
\end{tabular}

    \caption{Statistics for every drawing in the dataset of $(SGD)^2$. 
    Here $\sigma$ refers to the standard deviation of the statistic, indicating how much variance there is in the dataset.}
    \label{tab:rep_graph_statistics}
\end{table}

\begin{figure}[H]
\centering
\includegraphics[width=0.6\textwidth]{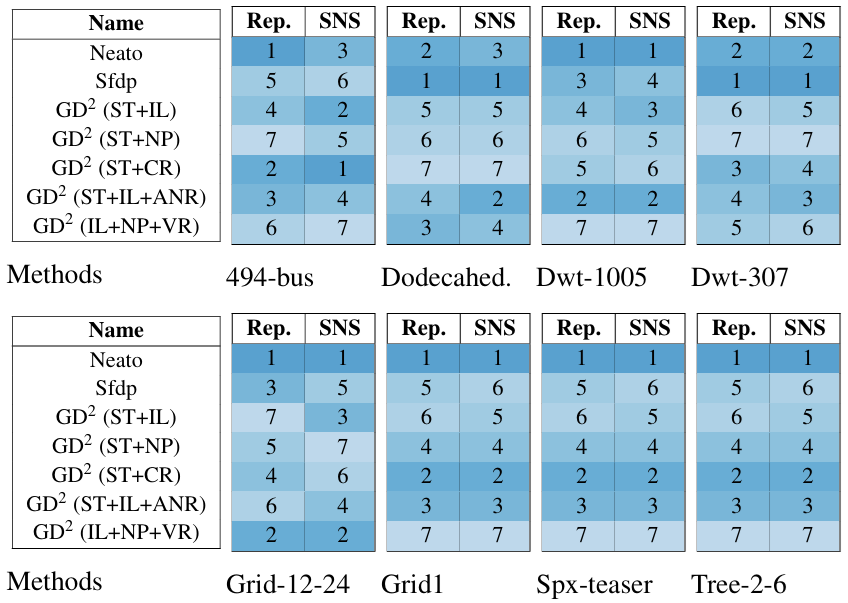}
\caption{The replicated ranking of results from~\cite{ahmed2022multicriteria} and the ranking of scale-normalized stress are compared. This comparison demonstrates that not only can the results change, but that there is a change for every dataset in the experiment.}
\label{tab:sgd-rep-sns}
\end{figure}

In our replication experiment, we replicate the environment of Ahmed et al.~\cite{ahmed2022multicriteria} and compute the normalized stress for a benchmark dataset of graphs using all evaluated algorithms. We recomputed the stress computations using scale-normalized stress. Note that the outputs (drawings) from the different algorithms were not modified -- the only difference is in the quality measure used: SNS instead of NS. The results are shown in~\autoref{tab:sgd-rep-sns}. 
A Jupyter notebook that reproduces the described experiment is included in our public repository. 

\subsubsection{Replication Results}
First of all (and not surprisingly) the values change when we switch from NS to SNS. The important question is whether the \textit{rankings} induced by these numbers (which determine what algorithm is considered better at preserving distances) change.

The replication experiment confirms that the rankings indeed change when we switch from NS to SNS.  
The proposed methods by Ahmed et al. are actually devalued when using \Normalizedstressmath, as their rankings improve when accounting for scale. The competing methods of NEATO and SFDP only lose rankings, see~\autoref{tab:sgd-rep-sns}.

The scale sensitivity of normalized stress is not an abstract problem -- it does occur in practice, and does affect results of empirical experiments. When computing stress, scale must be considered.

\begin{table}[ht]
    \centering
    \small 
    \begin{tabular}{|l|c c c c|c c c c|}
        \hline
        & \multicolumn{4}{c|}{\textbf{Rome-Lib Stats}} & \multicolumn{4}{c|}{\textbf{SS Stats}} \\ \hline
        \textbf{Statistic} & \textbf{Mean} & \pmb{$\sigma$} & \textbf{Min} & \textbf{Max} & \textbf{Mean} & \pmb{$\sigma$} & \textbf{Min} & \textbf{Max} \\ \hline
        $|V|$ & 46.776 & 28.014 & 10.0 & 100.0 & 331.179 & 277.232 & 3.0 & 1000.0 \\\hline
        $|E|$ & 46.198 & 33.675 & 10.0 & 100.0 & 402.853 & 366.533 & 3.0 & 1000.0 \\\hline
        Min Degree & 1.013 & 0.113 & 1.0 & 2.0 & 2.622 & 4.919 & 1.0 & 81.0 \\\hline
        Max Degree & 6.513 & 1.67 & 3.0 & 12.0 & 136.318 & 198.172 & 2.0 & 956.0 \\\hline
        Density $|E| / \binom{|V|}{2}$ & 0.083 & 0.057 & 0.025 & 0.4 & 0.068 & 0.126 & 0.002 & 1.0 \\\hline
        Diameter & 9.32 & 2.639 & 3.0 & 19.0 & 11.353 & 28.373 & 1.0 & 500.0 \\\hline
    \end{tabular}
    \caption{Statistics for every drawing in the two datasets: 1000 graphs from the Rome-Lib collection (left) and 1115 graphs from the SuiteSparse Matrix Collection (right). 
    Here $\sigma$ refers to the standard deviation of the statistic, indicating how much variance there is in the dataset.}
    \label{tab:graph_statistics}
\end{table}

\subsection{Validation Experiment}
\label{sec:experiments} 

\begin{table*}[ht]
    \centering
    {
    \begin{tabularx}{\textwidth}{|X|XX|XXX|} \hline
         & \textbf{\Rawstressmath} & \textbf{\Normalizedstressmath} & \textbf{\Minoptstressmath} & \textbf{\Shepardgoodnessmath} & \textbf{\Kruskalstressmath} \\\hline
        \textbf{\textit{$N<S<R$}} & \cellcolor[HTML]{FFFFFF} 0.00\% & \cellcolor[HTML]{FFFFFF} 0.00\% & \cellcolor[HTML]{7BB7DA} 96.10\% & \cellcolor[HTML]{97C6E2} 87.00\%  & \cellcolor[HTML]{85BCDD} 93.40\% \\\hline\hline
        $N<R<S$ & \cellcolor[HTML]{FFFFFF} 0.00\% & \cellcolor[HTML]{FFFFFF} 0.00\% & \cellcolor[HTML]{FFFFFF} 0.00\% & \cellcolor[HTML]{FFFFFF} 0.00\% & \cellcolor[HTML]{FFFFFF} 0.00\% \\\hline
       $ S<N<R $& \cellcolor[HTML]{FFFFFF} 0.00\% & \cellcolor[HTML]{FFFFFF} 0.00\% & \cellcolor[HTML]{EFF6FB} 3.90\% & \cellcolor[HTML]{D3E7F3} 13.00\% & \cellcolor[HTML]{E5F1F8} 6.60\% \\\hline
        $S<R<N$ & \cellcolor[HTML]{FFFFFF} 0.00\% & \cellcolor[HTML]{FFFFFF} 0.00\% & \cellcolor[HTML]{FFFFFF} 0.00\% & \cellcolor[HTML]{FFFFFF} 0.00\% & \cellcolor[HTML]{FFFFFF} 0.00\% \\\hline
        $R<N<S$ & \cellcolor[HTML]{9CC9E4} 84.70\% & \cellcolor[HTML]{9BC8E3} 85.30\% & \cellcolor[HTML]{FFFFFF} 0.00\% & \cellcolor[HTML]{FFFFFF} 0.00\% & \cellcolor[HTML]{FFFFFF} 0.00\% \\\hline
        $R<S<N$ & \cellcolor[HTML]{CEE4F1} 15.30\% & \cellcolor[HTML]{CFE5F2} 14.70\% & \cellcolor[HTML]{FFFFFF} 0.00\% & \cellcolor[HTML]{FFFFFF} 0.00\%  & \cellcolor[HTML]{FFFFFF} 0.00\% \\\hline
    \end{tabularx}
    }
    \medskip
    \caption{The relative frequencies of every 3-permutation of the three algorithms studied in this paper with respect to every metric, calculated on Dataset 1. Here N, S and R in the first column correspond to the drawing layouts NEATO, SFDP, and random, respectively. $X<Y$ indicates that $X$ has a better metric score than $Y$.}
    \label{tab:orders_analysis_rome}
\end{table*}

\begin{table*}[ht]
    \centering
    {
    \begin{tabularx}{\textwidth}{|X|XX|XXX|} \hline
         & \textbf{\Rawstressmath} & \textbf{\Normalizedstressmath} & \textbf{\Minoptstressmath} & \textbf{\Shepardgoodnessmath} & \textbf{\Kruskalstressmath} \\\hline
    \textbf{\textit{N $<$ S $<$ R}} & \cellcolor[HTML]{FFFFFF} 0.00\% & \cellcolor[HTML]{FFFFFF} 0.00\% & \cellcolor[HTML]{7DB8DB} 94.71\% & \cellcolor[HTML]{B3D5EA} 55.16\% & \cellcolor[HTML]{94C4E1} 85.65\% \\\hline\hline
    $N < R < S$ & \cellcolor[HTML]{FFFFFF} 0.00\% & \cellcolor[HTML]{FFFFFF} 0.00\% & \cellcolor[HTML]{FEFEFF} 0.27\% & \cellcolor[HTML]{FFFFFF} 0.09\% & \cellcolor[HTML]{FCFDFE} 0.90\% \\\hline
    $S < N < R$ & \cellcolor[HTML]{FFFFFF} 0.00\% & \cellcolor[HTML]{FFFFFF} 0.00\% & \cellcolor[HTML]{EEF6FA} 4.93\% & \cellcolor[HTML]{B7D8EB} 44.39\% & \cellcolor[HTML]{D9EAF4} 13.09\% \\\hline
    $S < R < N$ & \cellcolor[HTML]{FEFFFF} 0.18\% & \cellcolor[HTML]{FFFFFF} 0.00\% & \cellcolor[HTML]{FFFFFF} 0.00\% & \cellcolor[HTML]{FFFFFF} 0.00\% & \cellcolor[HTML]{FFFFFF} 0.00\% \\\hline
    $R < N < S$ & \cellcolor[HTML]{9DCAE4} 80.27\% & \cellcolor[HTML]{9DCAE4} 80.36\% & \cellcolor[HTML]{FFFFFF} 0.09\% & \cellcolor[HTML]{FEFFFF} 0.18\% & \cellcolor[HTML]{FEFEFF} 0.36\% \\\hline
    $R < S < N$ & \cellcolor[HTML]{CDE4F1} 19.55\% & \cellcolor[HTML]{CDE3F1} 19.64\% & \cellcolor[HTML]{FFFFFF} 0.00\% & \cellcolor[HTML]{FEFFFF} 0.18\% & \cellcolor[HTML]{FFFFFF} 0.00\% \\\hline
    \end{tabularx}
    }
    \medskip
    \caption{
    The relative frequencies of each potential ordering of the three algorithms studied in this paper with respect to each of the five metrics, calculated on Dataset 2. Here N, S and R in the first column correspond to the drawing layouts NEATO, SFDP, and random, respectively. $X<Y$ indicates that $X$ has a better metric score than $Y$. Notice how the scale-sensitive metrics of \Rawstressmath and \Normalizedstressmath consistently rank the random algorithm as the best. Meanwhile, \Minoptstressmath tends to recover our expected order.
    }
    \label{tab:orders_analysis_ss}
\end{table*}


While we have previously shown that not accounting for scale can effect quantitative evaluations, the results from our replication experiment are not dramatic. We now show through another experiment that not accounting for scale can  produce dramatically incorrect results.
We describe an empirical experiment intended to demonstrate that even a random layout can appear to have lower normalized stress than a stress-optimized graph layout. This does not occur in only a few strange edge cases, but can happen consistently when \normalizedstress is applied naively.

We conducted all experiments on a laptop  configured with Windows 11, 1.80 GHz AMD Ryzen 7 5700U, 16 GB RAM, and 8 logical cores. The code used was written in Python version 3.10. Below we describe the datasets and the graph layout algorithms employed in our experiments.

\subsubsection{Datasets}
We use subsets of two widely used graph collections as our dataset, chosen for their availability and thus reproducibility~\cite{di2024evaluating}. The range of graph sizes is consistent with state of the art evaluations~\cite{yoghourdjian2018exploring}.

\noindent
\textbf{Dataset 1 (Rome-Lib)}:
The first collection of graphs we use are from Rome-Lib~\cite{DiBattista1997b}, chosen for the variety and size of its graphs. This collection was obtained by performing random modifications such as vertex/edge addition/deletion from a set of 112 ``core'' real-world graphs, which had between 10-100 vertices. We randomly sampled a subset of 1,000 graphs from the full 11,534 graph collection. 

\noindent
\textbf{Dataset 2 (SuiteSparse Matrix Collection)}:
The second graph collection we make use of is the SuiteSparse Matrix Collection~\cite{davis2011university}, a standard benchmark for graph layout algorithms. 
SuiteSparse is an open source repository of real-valued matrices from different application areas, such as chemistry, infrastructure, and biology. Matrices can be interpreted as graphs by taking each row/column as a vertex and nonzero entries in the matrix as an edge. We use all graphs with up to 1,000 vertices which can be properly interpreted as a graphs, resulting in 1,115 graphs from the collection. The dataset statistics are shown in \autoref{tab:graph_statistics}.

Code to regenerate our graph datasets, along with statistics data, is available in our code \href{https://anonymous.4open.science/r/graph-layout-metrics-247D}{repository}.

\subsubsection{Graph Drawing Algorithms}
In this experiment, we applied six drawing algorithms to every graph in our two datasets.
From these six, we selected three algorithms which best demonstrate the discrepancy in scale: NEATO~\cite{gansner2005graph}, SFDP~\cite{ellson2001graphviz}, and a random drawing algorithm
(all algorithms are available at our code \href{https://anonymous.4open.science/r/graph-layout-metrics-247D}{repository}):
\begin{itemize}
    \item \textbf{NEATO} belongs to the MDS family of algorithms for graph layout. MDS explicitly minimizes the normalized stress of \autoref{eq:loss-stress}, using the computationally efficient majorization approach~\cite{gansner2005graph}. Hence, the stress score of the layout should be relatively small. However, NEATO outputs coordinates in pixel units.
    \item \textbf{SFDP} stands for Scalable Force-Directed Placement; it is a multi-level force-directed algorithm that can efficiently draw large graphs. It treats the nodes as particles in a simulation, which push and pull each other depending on their pairwise connectivity. A low energy configuration is found based on these forces~\cite{ellson2001graphviz}. While it does not explicitly optimize stress it produces aesthetically pleasing drawings that are associated with low stress~\cite{battista1998graph,eades2009graph}.
    \item \textbf{Random} layout is a random assignment of coordinates to the vertices within a unit square. Since it is randomly placing vertices, it is expected to produce drawings with no structure and high stress scores. As is typical in such settings, all vertex coordinates are in the unit square (in whatever internal coordinate representation).
\end{itemize}
We expect that with respect to stress the ``correct  order" of drawings obtained by these three algorithms should be NEATO, SFDP, random. NEATO directly minimizes stress, so it should clearly be superior in terms of distance preservation. SFDP does not directly utilize distances in its optimization, but it is doing something intelligent to untangle the edges of a graph and the output is generally readable. The random algorithm is completely divorced from the graph structure, so should never appear to preserve more information than the other algorithms.
However, NEATO and SFDP output much larger coordinates than the random algorithm.
As we will see, scale-invariant stress metrics indeed recover such orders, while the metrics affected by scale do not, as they are ``fooled" by the small size of the random algorithm.

\subsubsection{Metrics}
For every drawing of every graph in our corpus, we compute the two scale-sensitive values of \rawstress and \normalizedstress, as well as the three scale-invariant values of \minoptstress, \shepardgoodness, and \kruskalstress. We use the default scale of each drawing as the input to all scale-sensitive metrics. 
For each graph and for each metric we then compute the order of the drawings from best to worst (according to the metric under consideration).  
The full list of the five metrics that we evaluate is shown in ~\autoref{tab:stress_table}, and described in ~\autoref{sec:stress-defs}. 
All of the experimental data can be regenerated from the code in our repository.

\subsubsection{Validation Results and Discussion}

\textbf{Results}:
The results for the validation experiment on Dataset 1 are found in \autoref{tab:orders_analysis_rome}. 
Results from Dataset 2 are shown in ~\autoref{tab:orders_analysis_ss}. 
Most metrics agree that NEATO outperforms SFDP; however, the disagreement comes when we consider the random layout. Shockingly, scale-sensitive metrics consistently rank randomly drawn graphs as better than those obtained by NEATO or SFDP (99\% of the time)! Thus, the scale-sensitive metrics never achieve the ground-truth ordering. 

On the other hand, scale-invariant methods  determine the correct order with varying degrees of success, with \minoptstress achieving the expected ground-truth ordering 96.10\% of the time for Dataset 1, and 94.71\% for Dataset 2. 


The story of Dataset 1 is largely the same as that for Dataset 2, with the largest difference being that \Shepardgoodnessmath is much more accurate at 87\%. We believe this to be due to the fact that Dataset 1 contains much more sparser graphs in general, meaning that the NEATO algorithm has an easier time at capturing distances exactly with fewer tradeoffs.

\medskip\noindent
\textbf{Discussion}: \Normalizedstress always ranks the randomly drawn layouts better than those from NEATO and SFDP, which shows the potential for wrong results when not taking scale into account. \Rawstress also suffers from this problem. On the other hand, all scale-invariant metrics rank NEATO layouts as better than those from SFDP; and SFDP layouts as better than random layouts. Again, the only stress metric for graph layouts that consistently behaves as it should is \minoptstress. \Kruskalstress is ``right" more often, while \shepardgoodness is only correct around half of the time. When using raw stress or normalized stress, the scale of the output layouts holds a far greater impact on the quality metric than the actual quality of the layout. This is clearly wrong and justifies the need for scale-invariant quality metrics.


\subsection{Correlation Experiment}
We identify metrics that often agree on the ranking of drawings by computing the pairwise correlations between results of metrics. 

\textbf{Experiment Description and Results}:
Each graph-metric pair gives us some ordering of the selected algorithms. For each pair of metrics, we calculate the Spearman rank correlation coefficient based on the ordered metric scores across all graphs for the three layout algorithms. 
The correlations of all pairs of metrics are shown in Tables~\ref{tab:metric_pairs_correlations_rome}~and~\ref{tab:metric_pairs_correlations_ss}. Correlations have a range of $[-1,1]$, with a correlation score larger than zero indicating positive correlation and a negative correlation for scores of smaller than zero. We can conclude that pairs with high correlation agree on the order of best-to-worst in drawings more often than they disagree.

\begin{table}[ht]
    \centering
        {
        \small
        \begin{tabular}{|r |c c  | c c c|}\hline
         & \textbf{\Rawstressmath} & \textbf{\Normalizedstressmath} & \textbf{\Minoptstressmath} & \textbf{\Shepardgoodnessmath} & \textbf{\Kruskalstressmath} \\ \hline
        \RawStress & \cellcolor[HTML]{6BAED6} 1 & \cellcolor[HTML]{70B1D7} 0.989 & \cellcolor[HTML]{FED5B4} -0.367 & \cellcolor[HTML]{FED0AA} -0.416 & \cellcolor[HTML]{FED4B1} -0.382 \\\hline
        \NormalizedStress & \cellcolor{black!10} & \cellcolor[HTML]{6BAED6} 1 & \cellcolor[HTML]{FED5B3} -0.373 & \cellcolor[HTML]{FECFA8} -0.423 & \cellcolor[HTML]{FED3AF} -0.389 \\\hline
        \MinoptStress & \cellcolor{black!10} & \cellcolor{black!10} & \cellcolor[HTML]{6BAED6} 1 & \cellcolor[HTML]{7EB9DB} 0.954 & \cellcolor[HTML]{72B2D8} 0.984 \\\hline
        \ShepardGoodness & \cellcolor{black!10} & \cellcolor{black!10} & \cellcolor{black!10} & \cellcolor[HTML]{6BAED6} 1 & \cellcolor[HTML]{79B6DA} 0.967 \\\hline
        \KruskalStress & \cellcolor{black!10} & \cellcolor{black!10} & \cellcolor{black!10} & \cellcolor{black!10} & \cellcolor[HTML]{6BAED6} 1 \\\hline
        \end{tabular}
        }
    \medskip
    \caption{The Spearman correlations between various metrics when considering the ordering of the Random, SFDP, and NEATO drawing algorithms. These results are evaluated on a subset of 1,000 graphs from Dataset 1.}
    \label{tab:metric_pairs_correlations_rome}
\end{table}

\begin{table}[ht]
    \centering
        {
        \small
        \begin{tabular}{|r |c c  | c c c|}\hline
         & \textbf{\Rawstressmath} & \textbf{\Normalizedstressmath} & \textbf{\Minoptstressmath} & \textbf{\Shepardgoodnessmath} & \textbf{\Kruskalstressmath} \\ \hline
        \RawStress & \cellcolor[HTML]{6BAED6} 1 & \cellcolor[HTML]{6DAFD6} 0.994 & \cellcolor[HTML]{FEDABB} -0.33 & \cellcolor[HTML]{FDC392} -0.533 & \cellcolor[HTML]{FED5B3} -0.37 \\\hline
        \NormalizedStress & \cellcolor{black!10} & \cellcolor[HTML]{6BAED6} 1 & \cellcolor[HTML]{FEDABC} -0.329 & \cellcolor[HTML]{FDC392} -0.532 & \cellcolor[HTML]{FED5B3} -0.369 \\\hline
        \MinoptStress & \cellcolor{black!10} & \cellcolor{black!10} & \cellcolor[HTML]{6BAED6} 1 & \cellcolor[HTML]{A5CEE6} 0.779 & \cellcolor[HTML]{7EB9DB} 0.932 \\\hline
        \ShepardGoodness & \cellcolor{black!10} & \cellcolor{black!10} & \cellcolor{black!10} & \cellcolor[HTML]{6BAED6} 1 & \cellcolor[HTML]{A4CDE6} 0.786 \\\hline
        \KruskalStress & \cellcolor{black!10} & \cellcolor{black!10} & \cellcolor{black!10} & \cellcolor{black!10} & \cellcolor[HTML]{6BAED6} 1 \\\hline
        \end{tabular}
        }
    \medskip\caption{The Spearman correlations between various metrics when considering the ordering of the Random, SFDP, and NEATO drawing algorithms. These results are evaluated on a subset of 1,115 graphs from Dataset 2.}
    \label{tab:metric_pairs_correlations_ss}
\end{table}

Our results indicate that the scale-sensitive metrics (\rawstress and \normalizedstress) are highly positively correlated, with scores near 1. On the other hand the scale-invariant metrics (\minoptstress, \shepardgoodness, and \kruskalstress) are just as positively correlated among each other. However, between these groups we have strong negative correlations. However, each metric is trying to capture a similar concept. Between the scale-sensitive metrics, the correlation between \rawstress and \normalizedstress is 0.989 which is really large. 

The highest correlation between scale-invariant metrics is a value of 0.984 between \minoptstress and \kruskalstress. It is worth noting here that \minoptstress is the only one of the four variants of stress used to evaluate graph layouts that behaves as expected.

Even though Dataset 1 and Dataset 2 are different, the behavior of the eight metrics is similar (Table~\ref{tab:metric_pairs_correlations_rome} and Table~\ref{tab:metric_pairs_correlations_ss}). When going from the small graphs in Dataset 1 to the larger graphs in Dataset 2, the correlation between scale-sensitive metrics is unchanged, while the correlations between scale-invariant metrics are slightly weaker. On a high level, both datasets tell the same story: scale-sensitive metrics agree with each other, scale-invariant metrics also agree with each other, but there is no agreement between the two groups.

\medskip\noindent
\textbf{Discussion}:
If we look at the left blocks of Tables~\ref{tab:metric_pairs_correlations_rome}~and~\ref{tab:metric_pairs_correlations_ss}, we see that the scale-sensitive metrics are correlated. Which means that \rawstress and \normalizedstress agree quite often. This clearly indicates that the normalization factors in normalized stress are not behaving as expected, as there is little difference between using it and \rawstress. The correlations among scale-sensitive metrics are stronger than among scale-invariant metrics. This might be due to the fact that the equations of scale-sensitive metrics are more similar compared to the scale-invariant metrics. Our main takeaway is that there is a clear difference between the scale-sensitive and scale-invariant metrics: they agree within their category, but disagree between their categories.

\subsection{Stress Metric Running Times}
\label{sec:runtime_analysis}
We consider five quality evaluation metrics in this paper, and all achieve a quadratic complexity with respect to the number of vertices. These metrics compute a pairwise sum, straightforwardly giving a $O(|V|^2)$ complexity lower bound. 

\begin{figure}[ht]
    \centering
    \includegraphics[width=0.34\linewidth]{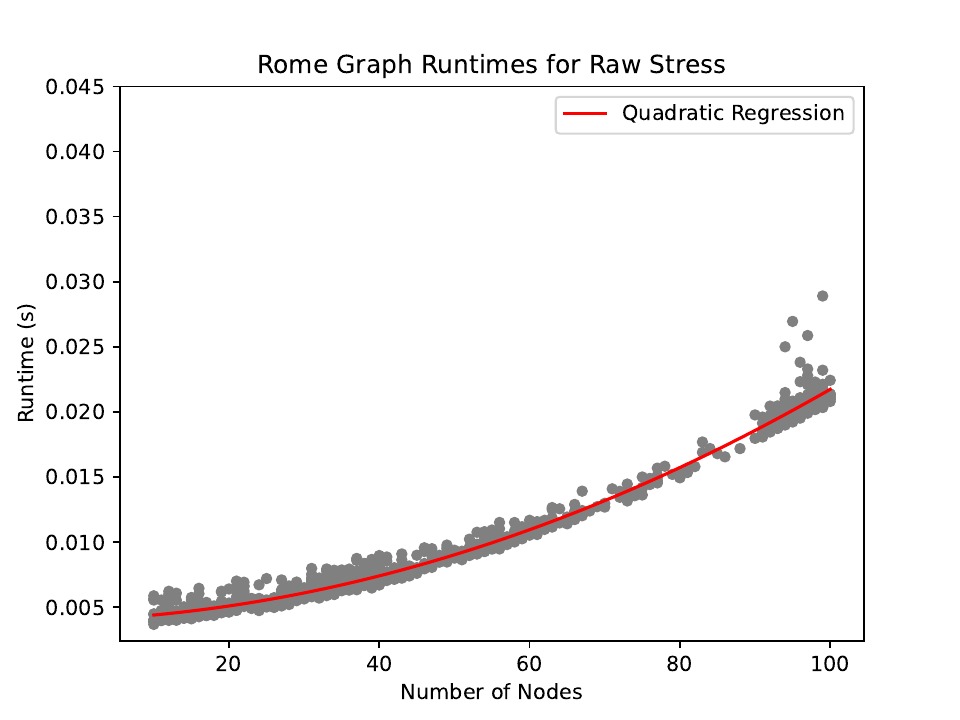}
    \includegraphics[width=0.34\linewidth]{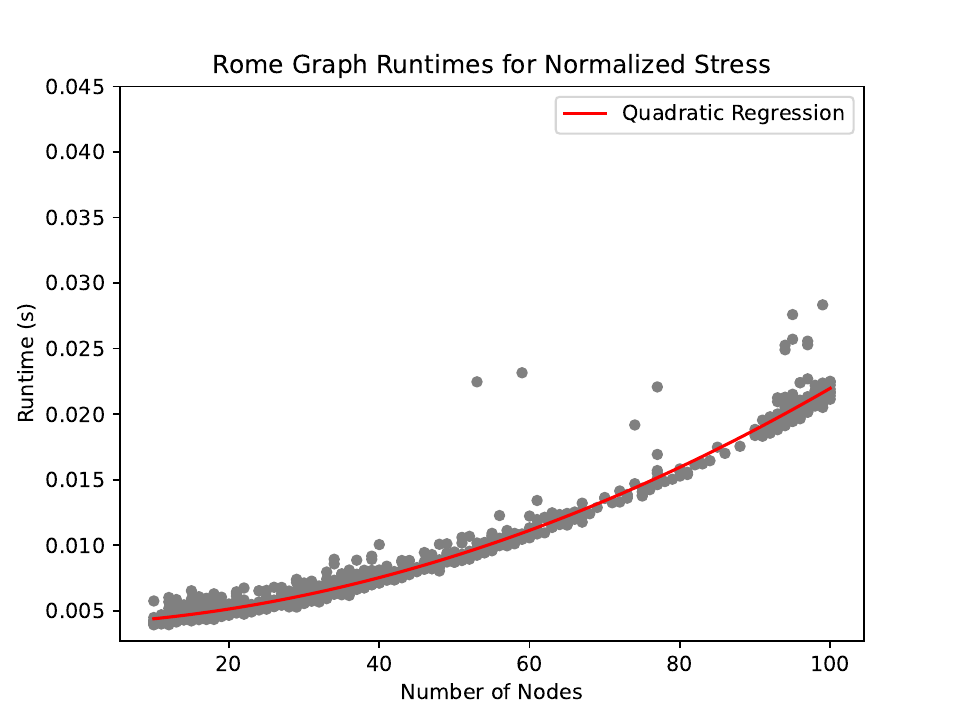}
    \includegraphics[width=0.34\linewidth]{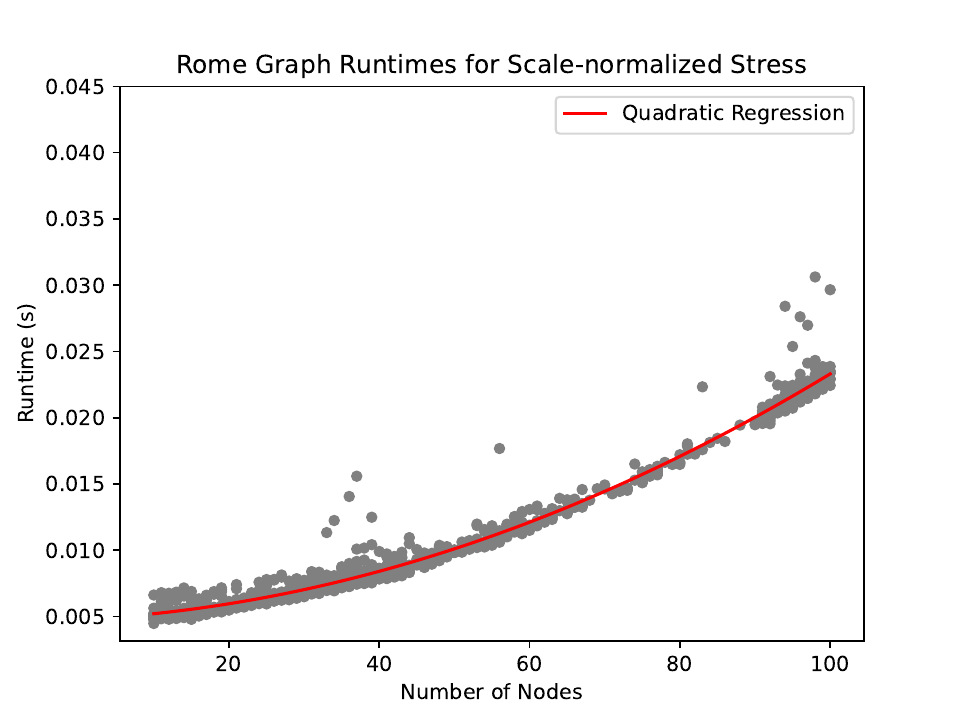}
    \includegraphics[width=0.34\linewidth]{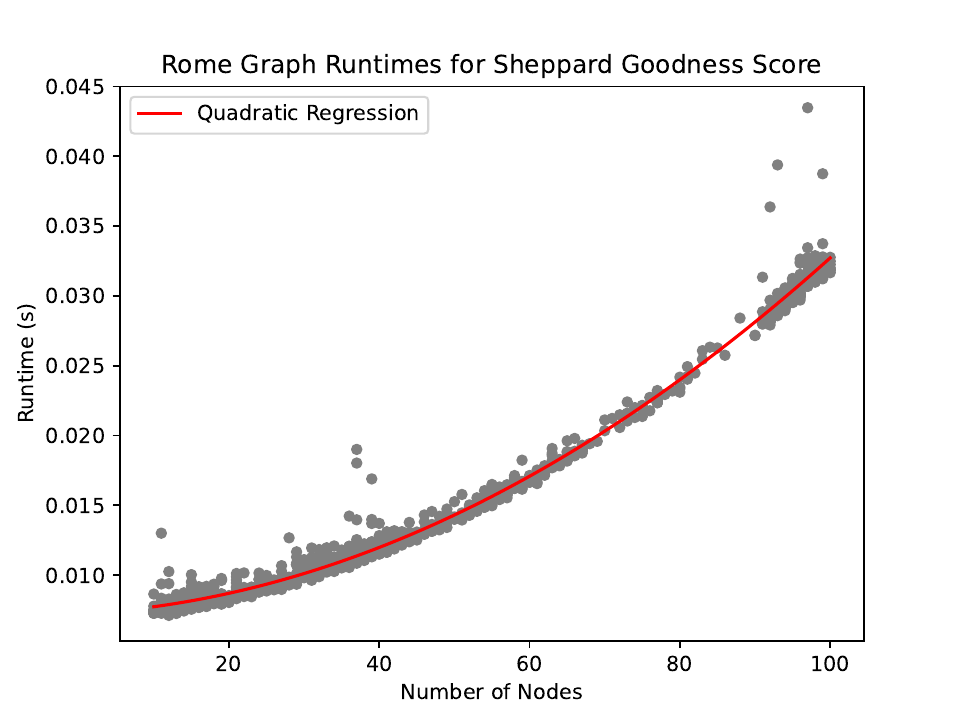}
    \includegraphics[width=0.34\linewidth]{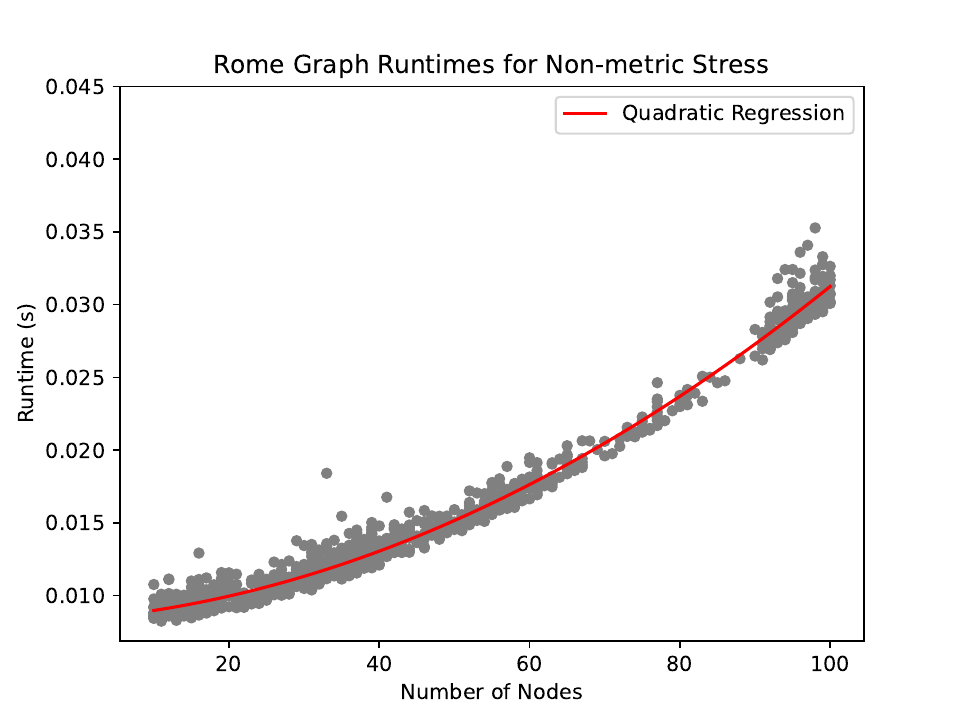}
    \caption{Experimental runtime analysis for the 1,000 graphs in dataset 1. Every point on the scatter plot maps the number of vertices in one graph to the total time in seconds required to compute the metric for six drawings. Quadratic regression lines are overlaid. 
    }
    \label{fig:runtime_analysis_rome}
\end{figure}

\begin{figure}[ht]
    \centering
    \includegraphics[width=0.34\linewidth]{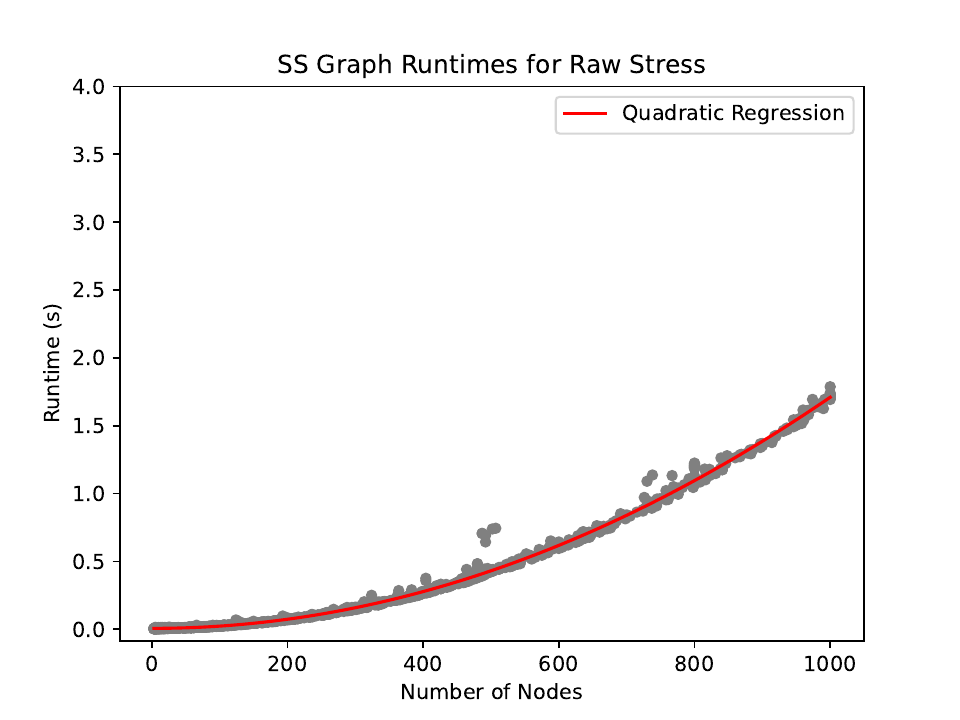}
    \includegraphics[width=0.34\linewidth]{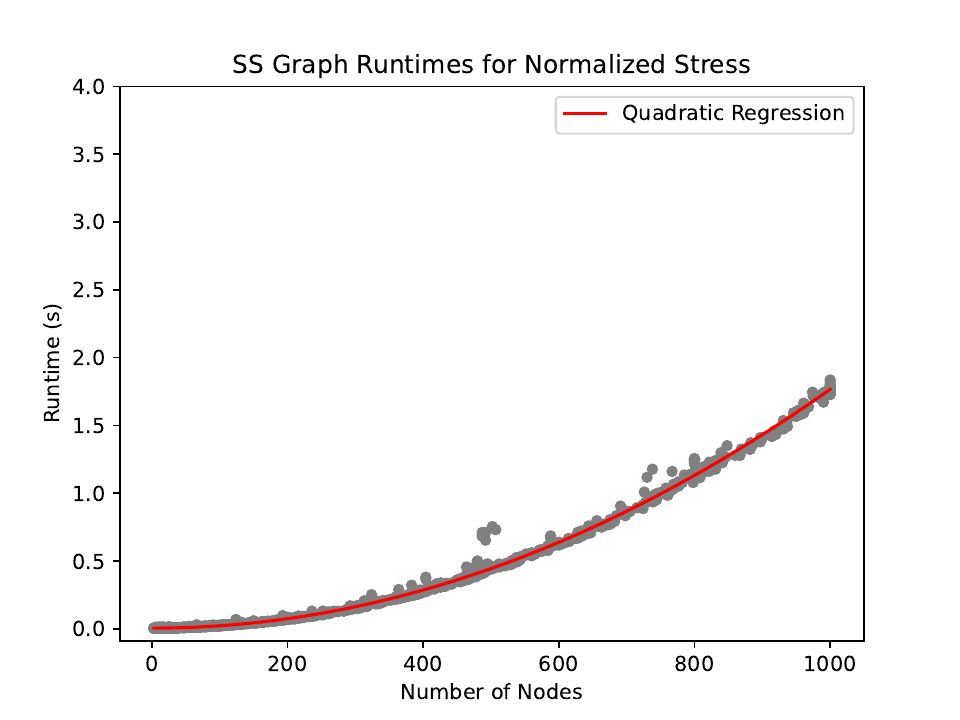}
    \includegraphics[width=0.34\linewidth]{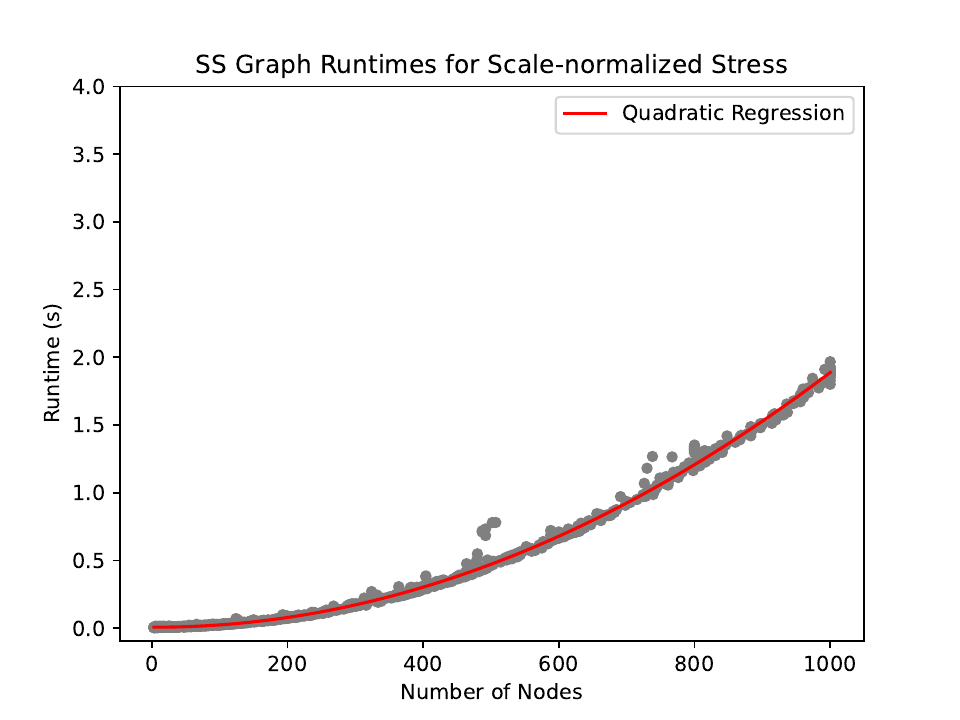}
    \includegraphics[width=0.34\linewidth]{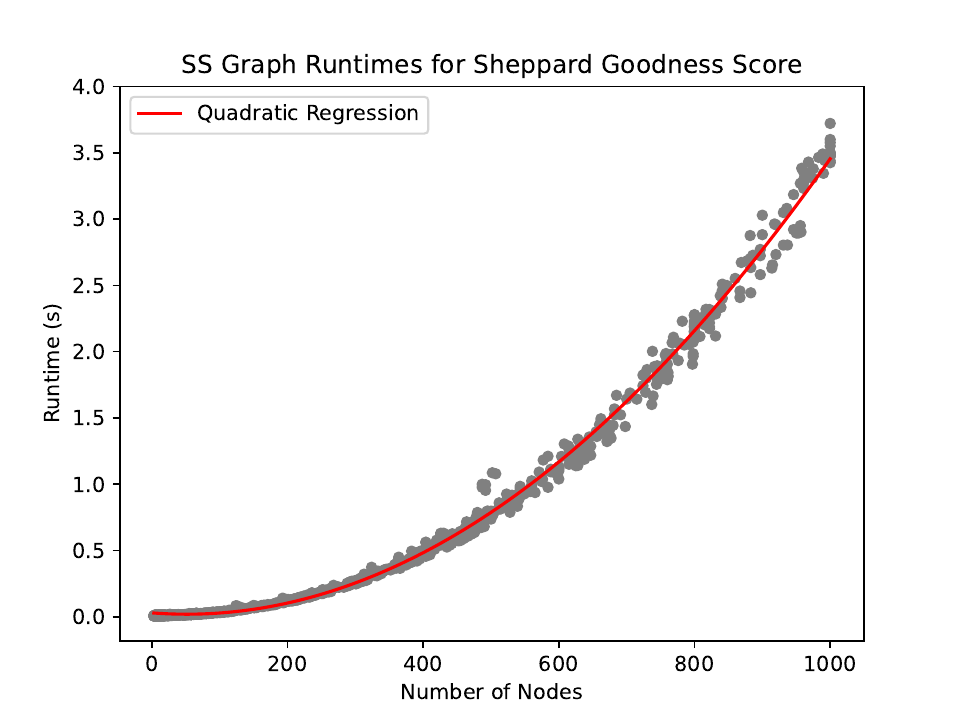}
    \includegraphics[width=0.34\linewidth]{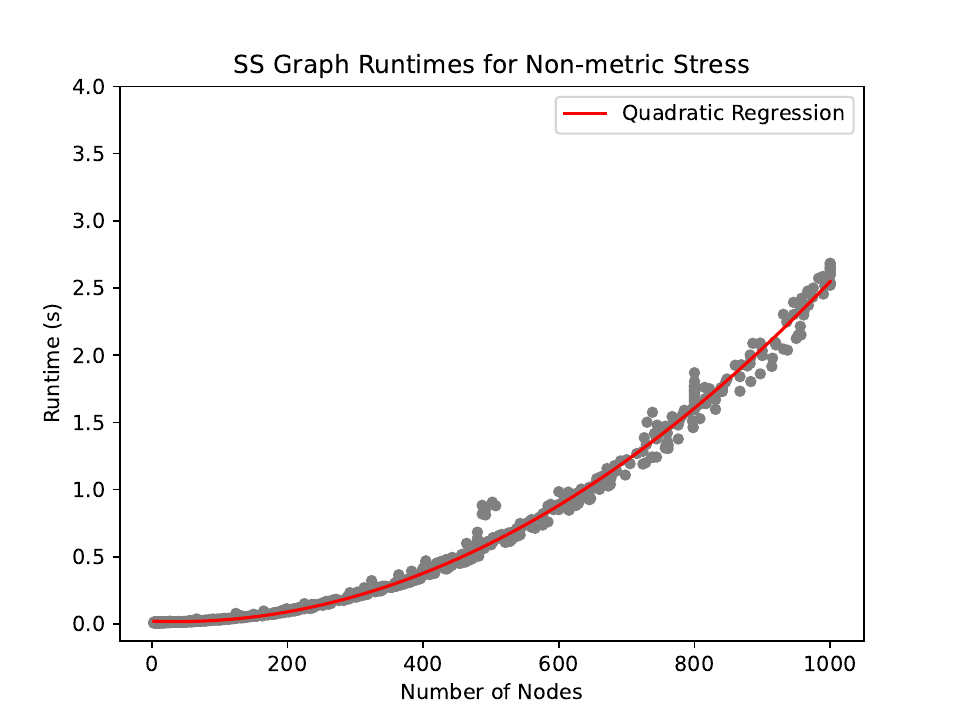}

    \caption{Experimental runtime analysis for the 1,115 graphs in Dataset 2. Every point on the scatter plot maps the number of vertices in one graph to the total time in seconds required to compute each metric for six drawings. Quadratic regression curves are shown overlaid on the results.
    }
    \label{fig:runtime_analysis_ss}

\end{figure}

In Figures~\ref{fig:runtime_analysis_rome}~and~\ref{fig:runtime_analysis_ss}, we see that the experimental running times of  \minoptstress, \normalizedstress, and \rawstress are comparable. The experimental running times of \shepardgoodness and \kruskalstress are slightly larger. Still, the running times are largely so similar that they should not be a factor in deciding which ones to employ for graphs with less than 1,000 nodes.


\subsection{Graph Layout Comparison}

\begin{figure}[htbp]
    \centering
    \includegraphics[width=\textwidth]{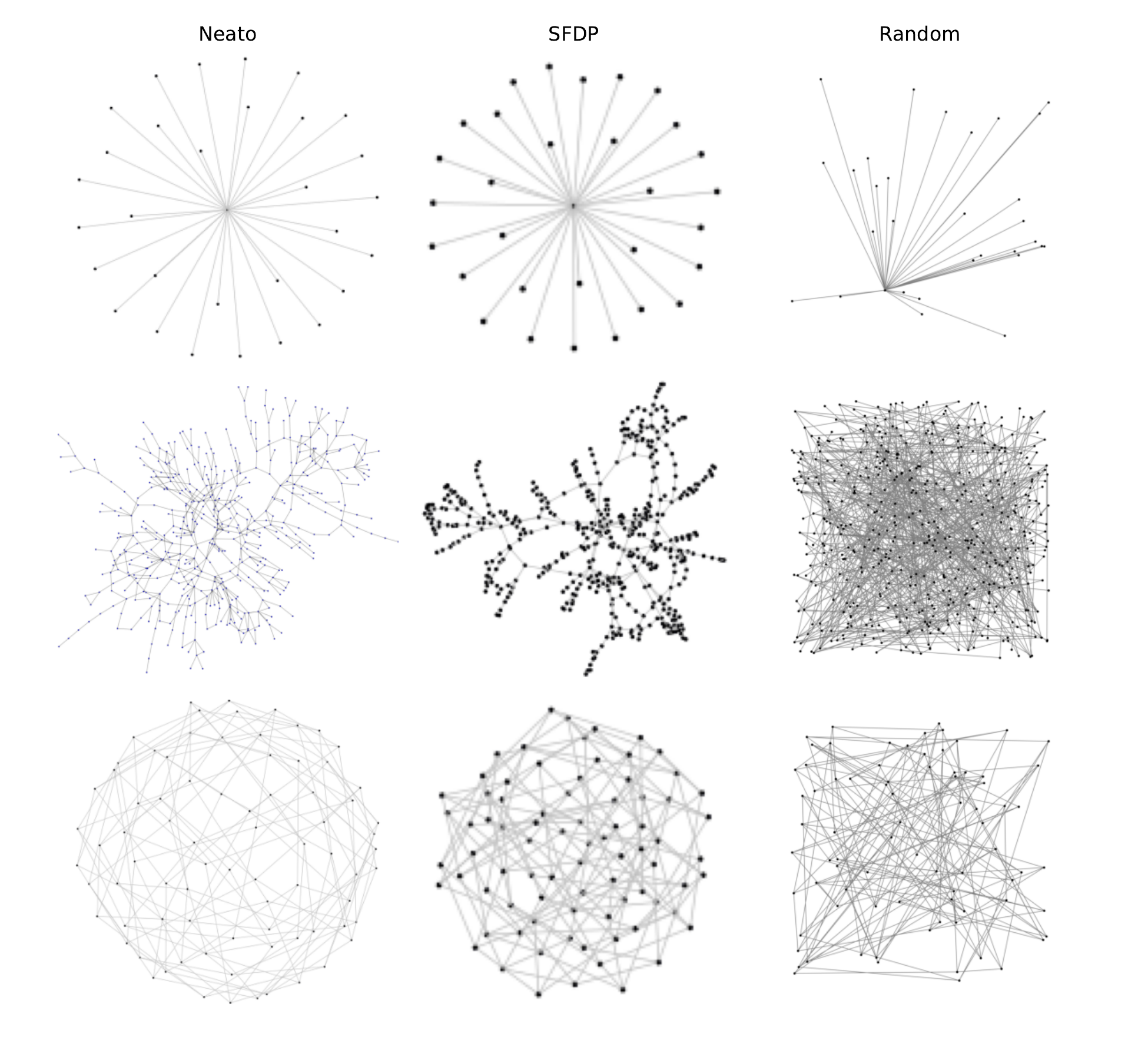}
    \caption{
        Layouts of three graphs from the SuiteSparse Matrix Collection visualized using the Neato, SFDP, and Random algorithms. Each row corresponds to a different graph (lp\_capri\_lo, 494\_bus, and cage\_G\_34), and each column shows the result of a different layout method.
    }
    \label{fig:suitesparse_layouts}
\end{figure}

\autoref{fig:suitesparse_layouts} illustrates the layouts of three representative graphs from the SuiteSparse Matrix Collection: \texttt{lp\_capri\_lo}, \texttt{494\_bus}, and \texttt{cage\_G\_34}. Each graph is visualized using three different layout algorithms: Neato, SFDP, and Random. The comparison highlights differences in visual clarity and structural insight provided by each algorithm. Neato and SFDP generally preserve global structure better than the Random layout, which serves as a baseline for spatial distribution without structural optimization.

\section{Limitations}
In our experiment, we evaluated five distinct stress metrics, on two datasets, using three layout algorithms: NEATO, SFDP, and Random. Dataset 1 and Dataset 2  used in our experiment are relatively small and sparse. Replicating the experiment on larger and more diverse graphs will improve the generalizability of the results reported in this paper.
There are many more graph layout algorithms that we could have considered (and even more $k$-tuples for the order comparisons). Replicating the experiments with popular layout algorithms such as  ForceAtlas2~\cite{jacomy2014forceatlas2}, Graphviz's Dot~\cite{ellson2001graphviz}, and GEM~\cite{frick1995fast} would also help with generalizability. While we attempted to cover the spectrum of stress-based metrics used in graph drawing and dimensionality reduction, there are more variants than the five used here. 

Our experimental validation of the five metrics under consideration was based on selecting three layout algorithms and the expectation that there is a ``correct order" when considering how well they capture the graph distances in their embeddings. It is likely that a better-grounded experiment could be devised to validate the efficacy of stress-based metrics.

\section{Conclusions and Future Work}
We have shown that comparing scale-sensitive stress values can lead to clearly incorrect conclusions. We have reviewed publications that use or refer to stress, showing inconsistencies and misinterpretations. We also analytically derived several properties of the most popular version -- normalized stress -- and demonstrated its scale sensitivity, to give readers an intuition about why this fails. We conducted two numerical experiments, the first to show that scale-sensitive measures have affected previous evaluations, and the second to show that even a completely random layout can consistently outperform a stress-minimized layout when scale is not taken into account.

Of the three scale-invariant measures we considered, the \kruskalstress appears to be a promising candidate for a normalized, 0-1, scale-invariant stress metric. However, it is among the least explainable metrics we investigated, and it achieved only 85\% accuracy on dataset 2. Further investigation should be done on this metric. The Shepard goodness score also shares many of these properties, and is very interpretable, but appears to give us mixed results on our expected ranking. 

The ``gold standard" we recommend for graph layout researchers is the scale-normalized stress measure. It is the simplest of the scale-invariant metrics to explain and has been employed in several previous evaluations.



Our recommendations are based on a carefully designed experiment that provides a natural expected ordering of the quality of different layouts for any stress-based quality metric. The experiment uses two frequently studied graph datasets: Rome-Lib and the SuiteSparse Matrix Collection. Our experimental results imply that scale-sensitive metrics, although often used in practice, are not able to accurately compare layouts obtained by different algorithms. On the other hand, scale-invariant metrics often provide the expected ordering. All metric implementations, experimental data and analysis are available online\footnote{\small \url{https://anonymous.4open.science/r/graph-layout-metrics-247D}}.

The results in this paper call into question the validity of evaluations that make use of scale-sensitive metrics. Other graph drawing metrics (e.g., symmetries~\cite{welch2017measuring}) are also affected by scale. Careful consideration of scale-invariant metrics can help develop better layout algorithms and more convincing comparisons between different layout algorithms.
It remains as future work to re-evaluate prior studies using  scale-invariant metrics in order to verify their results. 

{\color{blue} 

}

\bibliography{references, survey }



\end{document}